\documentclass[11pt,manuscript,times]{aastex62}

\usepackage{amsmath,amssymb,amstext}
\usepackage{times}
\usepackage{tabularx}
\usepackage{longtable}
\usepackage{float}
\usepackage{natbib}

\newcommand{\ha}{H$\alpha$}
\newcommand{\comments}[1]{}

\begin{document}


\title{MAGNETIC FIELD DYNAMICS AND VARYING PLASMA EMISSION IN LARGE SCALE CORONAL LOOPS}

\author{\c{S}ahin, S}
\affil{\it Akdeniz University, Faculty of Science, Department of Space Science and Technologies, 07058, Antalya, Turkey}

\author{Yurchyshyn, V.}
\affil{\it Big Bear Solar Observatory, New Jersey Institute of Technology, Big Bear City, CA 92314, USA}

\author{Kumar, P.}
\affil{\it Heliophysics Science Division, NASA Goddard Space Flight Center, Greenbelt, MD 20771, USA}

\author{Kilcik, A}
\affil{\it Akdeniz University, Faculty of Science, Department of Space Science and Technologies, 07058, Antalya, Turkey}

\author{Ahn, K}
\affil{\it Big Bear Solar Observatory, New Jersey Institute of Technology, Big Bear City, CA 92314, USA}

\author{Yang, X.}
\affil{\it Big Bear Solar Observatory, New Jersey Institute of Technology, Big Bear City, CA 92314, USA}

\begin{abstract}

In this study we report detailed observations of magnetic environment at four footpoints of two warm coronal loops observed on 5 May 2016 in NOAA AR 12542 (Loop I) and 17 Dec 2015 in NOAA AR 12470 (Loop II). These loops were connecting a plage region with sunspot periphery (Loop I) and a sunspot umbra (Loop II). We used Solar Dynamics Observatory (SDO) and Goode Solar Telescope (GST) data to describe the phenomenon and understand its causes. The study indicates loop brightening episodes were associated with magnetic flux emergence and cancellation processes observed in SDO's Helioseismic and Magnetic Imager (HMI) and GST's Near InfraRed Imaging Spectrapolarimeter (NIRIS) data. The observed activity was driven by magnetic reconnection between small-scale emerging dipoles and large-scale pre-existing fields, suggesting that the reconnection occurred in the lower chromosphere at the edge of an extended plage region, where the loops were rooted. We suggest that plasma, evaporated during these reconnection events, gradually filled the loops and as it cooled the visible density front propagated from one footpoint of the loop to another at a rate of 90-110 km s$^{-1}$. This study also indicates that at least some of the bright loops seen in SDO Atmospheric Imaging Assembly images rooted in sunspot umbra may be heated due to magnetic activity taking place at the remote (non-sunspot) footpoint.

\end{abstract}

\comments{Key words: Sun: atmosphere -- Sun: chromosphere -- Sun: flares}

\section{Introduction}

\noindent  Coronal loops are observed in UV and X-ray images as bright curved arches that can extend to a quite large fraction of the solar radius. They are thought to be hot and dense plasma confined by guiding magnetic flux tubes \citep[e.g.,][]{2004A&A...428..629M}. Coronal loops can be separated into three different groups depending on their plasma temperatures: i) cool loops (0.1-1 MK), which were first detected in Ultra Violet (UV) lines by \cite{1976ApJ...210..575F}, ii) warm loops (1-2.0 MK) that are well observed in extreme UV (EUV) \citep{1999ApJ...517..497L,2003A&A...406L...5D}, and iii) hot loops ($>$ 2MK), which comprise most of the structures visible in X-ray images \citep{2003ApJ...590.1095N}. The electron density of coronal loops are measured to range from 10$^{8}$ up to 10$^{12}$~cm$^{-3}$, with the highest values typically only seen in flaring loops \citep{2010LRSP....7....5R}.

Plasma diagnostics such as Doppler velocity, density, and temperature measurements are of particular interest as they reveal the thermal evolution of coronal loops needed for identification of the heating mechanism. For a recent review on various aspects of coronal loops and 3D modeling see \cite{Reale2014} and \cite{2015RSPTA.37350055P}. Imaging and spectroscopic observations in combination with spectrapolarimetry data have provided information on the coupling between plasma flows and guiding coronal fields \citep[e.g.,][]{2004A&A...428..629M} and have improved our understanding of physical conditions in coronal loops \citep[e.g.,][]{2015ApJ...800..140G} and flows within these loop structures \citep[e.g.,][]{2003A&A...406.1089D}.

Warm coronal loops comprise the majority of intensity structures seen in active regions (ARs) in the UV spectral range. Their footpoints are emitting at temperatures of around 1.0 MK and their electron density is of order of 10$^{9}$~cm$^{-3}$ \citep{2003A&A...406L...5D}. They often show downflows in transition region (TR) lines and upflows in UV lines \citep{2009ApJ...694.1256T} as well as notable non-thermal velocities near the footpoints \citep{2008ApJ...678L..67H,2012ApJ...759..104H}. In particular, \cite{2009ApJ...694.1256T} reported that upflows in coronal loops increase with higher plasma temperatures, while downflows near the loop footpoints may reach up to 60~km s$^{-1}$, when observed at lower temperature lines. \cite{2002ApJ...567L..89W} detected line-of-sight (LOS) flows along warm loops of up to 40~km s$^{-1}$, while \cite{2008A&A...482L...9O} reported speeds of 74 - 123 km s$^{-1}$.

The thermal structure of coronal loops is being extensively studied since it may hint to us as to what mechanism may be responsible for coronal heating. High resolution UV observations seem to indicate that at spatial scales below 365~km coronal loops exhibit constant density and temperature across their width \citep{2003A&A...406.1089D}. \cite{2014ApJ...787L..10W} used High resolution Coronal Imager (Hi-C) data to find that 70\% of Hi-C loop pixels do not show evidence for existence of thermal substructures, while \cite{0004-637X-840-1-4} concluded that the Hi-C instrument fully resolves the structure of coronal loops. \cite{0004-637X-840-1-4} further argued that the model derived minimal loop width of 550~km is defined by the spatial extent of the corresponding energy release events, which should be in this case of granular (macroscopic) scales, thus ruling out the magnetic field braiding mechanism operating on unresolved microscopic scales. At the same time \cite{2017ApJ...837..108P} simulations revealed that braided structures, even if they are present in the magnetic field, may not always be readily visible in EUV images as such.

Although progress has been made studying physical properties of coronal loops, accurate knowledge of their heating mechanism still eludes us. It is commonly accepted that energy for coronal heating is stored in the magnetic field and is converted from mechanic energy of convective motions of solar plasma that stretch, displace, twist, and braid magnetic field lines \citep{2006SoPh..234...41K}. The stored magnetic energy may then be released either via magnetic reconnection process (DC, direct current heating) or dissipation of waves (AC, alternating current heating). Reconnection requires either the existence of mixed polarity fields at the loop footpoints (macroscopics events) or highly braided loops along their length \citep[microscopic scales,][]{2013Natur.493..501C}. Until recently, both conditions were not frequently observed in the solar atmosphere.

The AC heating mechanism \citep{2006SoPh..234...41K} may be realized due to rigorous turbulent flows in the photosphere \citep{Abramenko_2011,2013ApJ...773..167A} capable of rapidly displacing loop footpoints over time scales shorter than those required for Alfv{\' e}n wave to travel from one footpoint to another. \cite{2014A&A...564A..12C} simulations showed that bright UV loops in an emerging active region (AR) may be formed due to dissipation of electric currents induced by advection of small-scale magnetic elements. Recently \cite{0004-637X-849-1-46} introduced a coronal heating model by Alfv{\' e}n wave turbulence, which predicted that neither short period waves launched within flux tubes nor long-period random footpoint motions are able to satisfactory reproduce the observed physical parameters and energy budget. \cite{2018arXiv180206206Y} modelled loop-loop reconnection to argue that this mechanism is capable of providing energy necessary for coronal heating. Since the simulated interacting loops were of comparable spatial scale, this numerical experiment argues in support of coronal heating via loop braiding mechanism \citep{1988ApJ...330..474P,2006SoPh..234...41K, 2017ApJ...837..108P}.

In spite of a decades long effort little is known about the magnetic configuration at footpoints of coronal loops. With the advance of solar instrumentation it becomes evident that coronal loops are often rooted in a mixed polarity and dynamic fields this is in stark contrast to the long prevailing view that plage fields in solar ARs are mainly unipolar. Thus, \cite{2012ApJ...760...82S} reported that high-speed outflows were only observed in association with mixed polarity fields observed with Solar Dynamics Observatory's (SDO) Helioseismic and Magnetic Imager (HMI) instrument. \cite{2017ApJS..229....4C,2018A&A...615L...9C} reported that studied coronal loops were rooted in a small-scale, mixed polarity magnetic environment and argued that flux cancellation and reconnection low in the solar atmosphere drive mass and energy flows along the loops. \cite{2014A&A...564A..12C}'s simulations further detail the process on interaction between convection and emerging magnetic flux that may lead to enhanced heating at footpoints of coronal loops. Earlier, \cite{2003ApJ...593..549F} proposed that heating of coronal loops observed in quiet Sun regions may be fueled by ``explosions'' of granule-scale sheared magnetic bipoles emerging at the edge of network flux concentrations. Using non linear force-free field extrapolation, \cite{2017ApJ...843L..20T} found that those loops connecting to plage regions, penumbra of opposite polarity sunspot, or to a mixed-polarity flux region are the brightest loops, while the umbra-to-umbra loops remain mostly invisible. These findings further emphasize and support the idea that magnetoconvection and magnetic field cancellation may play a vital role in coronal heating.

The UV and X-ray sunspot rooted loops \citep{1976ApJ...210..575F} are of particular interest since it is not clear in this case what mechanism may be responsible for plasma heating and acceleration. Early observations of umbral loops were mostly performed in X-ray spectral range and therefore they possibly only address hot loops as defined above, while most of the observed sunspot loops are now classified as warm and they appear in UV images. Thus, \cite{1992ApJ...399..313S} used a limited size X-ray data set to argue that no bright loops were detected rooted in the umbra. \cite{Webb1981} reported that ``no non-flaring X-ray loops end in umbra''. \cite{2005ApJ...621..498K} showed that the cool loops, mostly found rooted in pores and sunspots are associated with high magnetic filling factor suggesting that lack of heating could be due to suppressed magneto-convection in the strong-field umbral regions.

\cite{2007ASPC..369..287K} examined footpoint locations using continuum intensity data and found that about half of the loops were anchored at the umbra-penumbra boundary region, while nearly equal parts of the remaining loops were located in umbra and penumbra with the tendency for brighter loops to be predominantly rooted in umbra. Authors argued that irregularities in the sunspot magnetic field introduced by light bridges (LBs) or sunspot fragmentation \citep[e.g.,][]{2014A&A...568A..60L,Tian_2014,2015ApJ...798..136Y} lead to formation of current sheets resulting in coronal heating. \cite{2016A&A...587A..20C} analyzed slit-jaw images and spectroscopic data from the Interface Region Imaging Spectrograph \citep[IRIS,][]{iris} and did not detect any direct evidence of energy input at a footpoint of a bright coronal loop rooted at a sunspot. Earlier reports \citep[e.g.,][]{2003A&A...406L...5D,2009ASPC..415..241U} similarly suggested that activity at other , non-sunspot footpoint of the loop may be the cause of their enhanced temperature and density. 

In this study we focus our effort on footpoints of two warm coronal loops well observed in SDO's Atmospheric Imaging Assembly (AIA) 171~\AA\ images. Both loops had their one footpoint rooted in (Loop II) or nearby (Loop I) of a sunspot, while the other one (``remote'') located in an AR plage. Our goal is to study the dynamic of the underlying magnetic field co-spatial with the loop footpoints and co-temporal with episodes of loop brightening and to determine their possible role in the observed loop brightening. One of the known difficulties related to coronal loop studies is contamination of loop emission by overlapping emission from other loops and bright UV background as well as insufficient spatial resolution that often prevents reliable identification of loop footpoints and affects loops diagnostics. It was therefore our objective to select isolated loops that can be reliably traced from one footpoint to another. We also take advantage of high resolution measurements of the photospheric magnetic field provided by the Goode Solar Telescope (GST). The loop observed on 10 May 2016 (Loop I) was connecting an inner part of NOAA AR 12542 occupied by pores with a plage area east of the AR. We analyzed one loop revival episode that took place between 17:00~UT and 17:30~UT. The loop on 12 Dec 2015 (Loop II) was also observed between 17:00~UT and 17:30~UT in an extended NOAA AR 12470 connecting the main leading sunspot with a peripheral plage area east of the sunspot. Both loops were large-scale with the footpoint separation of 172$\arcsec$ and 140$\arcsec$, respectively, and their width fluctuated between 1 and 2~Mm. In Section 2 we describe observational data and in Section 3 we present results for two coronal loops. Conclusions and Discussion are in Section 3.

\section{Data}

\noindent 
The data in this study were collected using GST TiO broadband imager, Visible Imaging Spectrometer \citep[VIS,][]{Cao2010} and Near-Infra-Red Imaging Spectropolarimeter \citep[NIRIS,][]{niris} as well as Helioseismic and Magnetic Imager \citep[HMI,][]{hmi,hmi2} and the Atmospheric Imaging Assembly \citep[AIA,][]{aia} instruments on board of Solar Dynamics Observatory \citep[SDO,][]{sdo}.

Photospheric TiO images were acquired every 15~s using a 10~\AA\ bandpass TiO filter centered at 7057~\AA\ with the pixel scale of 0$\arcsec$.0375. The VIS combines a 5~\AA\ interference pre-filter with a Fabry-P\'{e}rot etalon to produce a band pass of 0.07~\AA\ over a round 70$\arcsec$ wide field of view (FOV). The pixel scale is 0$\arcsec$.029. The difference in the acquisition time at two sequential line positions (e.g, +0.8~\AA\ and -0.8~\AA) was about 2~s. All images were acquired with the aid of an adaptive optics (AO) system, which incorporates a 357 actuator deformable mirror, a Shack-Hartmann wavefront sensor with 308 sub-apertures, and a digital signal processor system \citep{2014SPIE.9148E..50Z}. All TiO and VIS data were speckle reconstructed with the Kiepenheuer-Institut f{\"u}r Sonnenphysik's software package \citep[KISIP,][]{kisip_code} to achieve the diffraction limit of the telescope (0$\arcsec$.1) over a large FOV. 

The NIRIS vector magnetic field data was acquired using AO corrected light, a dual Fabry-P\'erot etalon and a 2k $\times$ 2k HgCdTe Helium cooled Teledyne camera. Two polarization states are simultaneously imaged side-by-side on a 1024 $\times$ 1024 pixel area each, using a dual beam system that provides a 85$\arcsec$ round FOV with image scale of 0$\arcsec$.083/pixel. The measurements were performed using the Fe I 15650~\AA\ doublet with a bandpass of 0.1~\AA\ and a rotating 0.35$\lambda$ wave plate that allowed us to sample 16 phase angles at each of more than 60 line positions at a cadence of 30 s per one full spectroscopic measurement (full-Stokes I, Q, U, V). The Fe I 15650~\AA\ Stokes data were corrected for polarization effect and inverted using a Milne-Eddington (ME) inversion approach adopted for NIRIS data. This inversion code was written by J. Chae (private communication) using the fomulae given in \cite{1992soti.book...71L}. Its early version was previously applied to the Hinode/SP data by \cite{2009ASPC..415..101C}. The code sets the filling factor/stray light fraction parameter to unity, which is because magnetic structures are believed to be fully resolved in these data. An inverted data set includes nine parameters among which are the total magnetic field strength, inclination and azimuth angles, and the Doppler shift. For ME code performance comparison, see \cite{2014A&A...572A..54B}.

The SDO/AIA instrument acquires full disk EUV images of the Sun (FOV $\sim$1.3 R$_\odot$) with a spatial resolution of 1$\arcsec$.5 (0$\arcsec$.6 per pixel) and a cadence of 12~s. In this study, we relayed on 171 \AA\ (Fe IX, $T\approx$0.7 MK) data since these images were less populated by loops at various temperatures and densities which allowed us to reliably identify loop footpoints. To determine coronal temperatures and emission measure (EM) we utilized data from six AIA channels: 94~\AA\ (Fe XVIII, $T\approx$6.3~MK), 131~\AA\ (Fe VIII, Fe XXI, Fe XXIII, i.e., 0.4, 10, 16~MK), 171~\AA\ (Fe IX,$T\approx$0.7~MK), 211~\AA\ (Fe XIV, $T\approx$2~MK), 193~\AA\ (Fe XII, Fe XXIV, $T\approx$1.5, 20~MK) and 335~\AA\ (Fe XVI, $T\approx$2.5 MK). To analyse magnetic fields at the footpoints we also used HMI {\it hmi.B\_720s} series data, which are HMI full-disk VFISV ME inverted and disambiguated vector field observations \citep{hmidata,hmidata2}.

\section{Results}

\subsection{10 May 2016 Loop}

\begin{figure}
\centering{
\includegraphics[width=3.2truein]{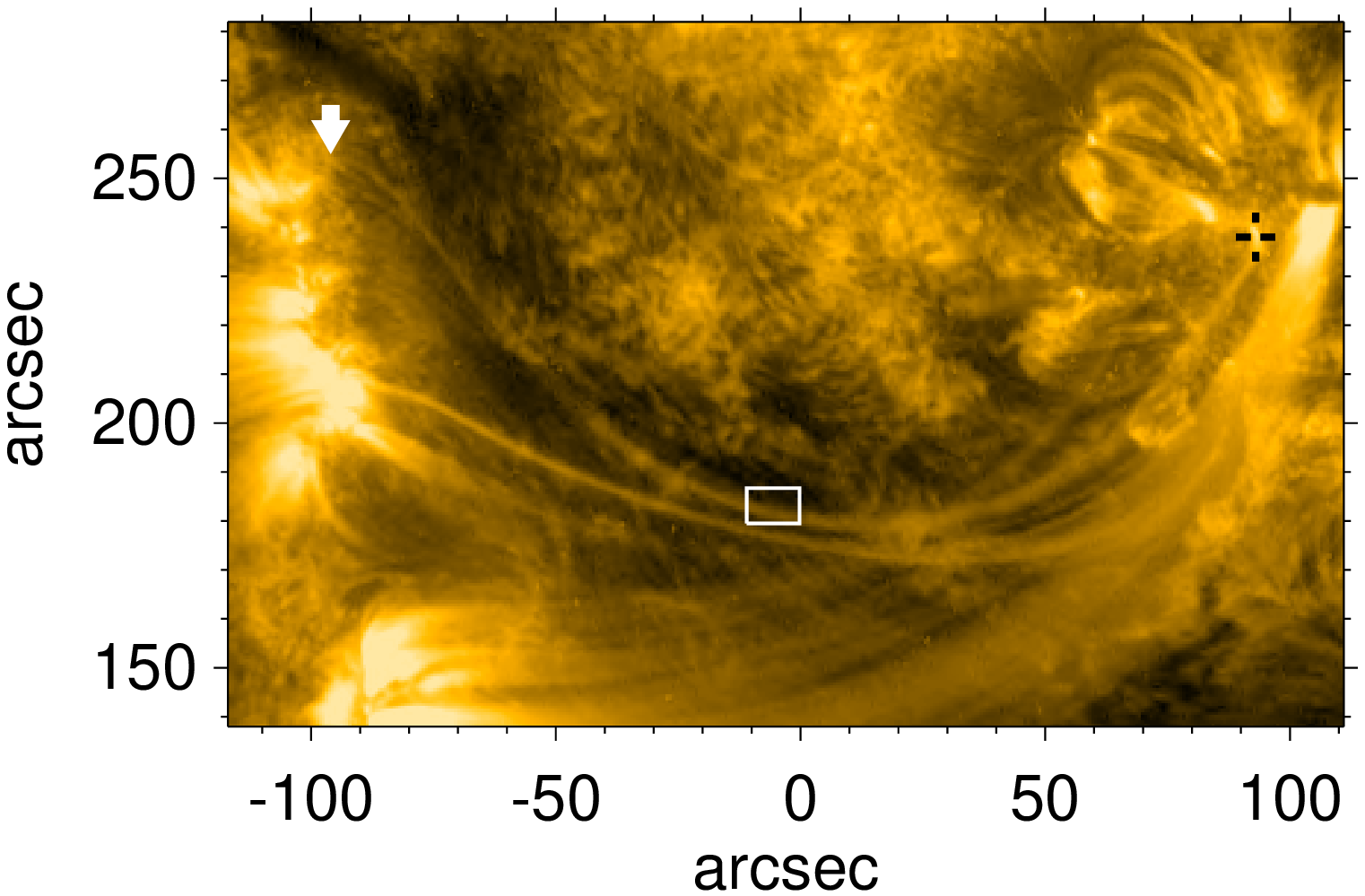}
\includegraphics[width=3.2truein]{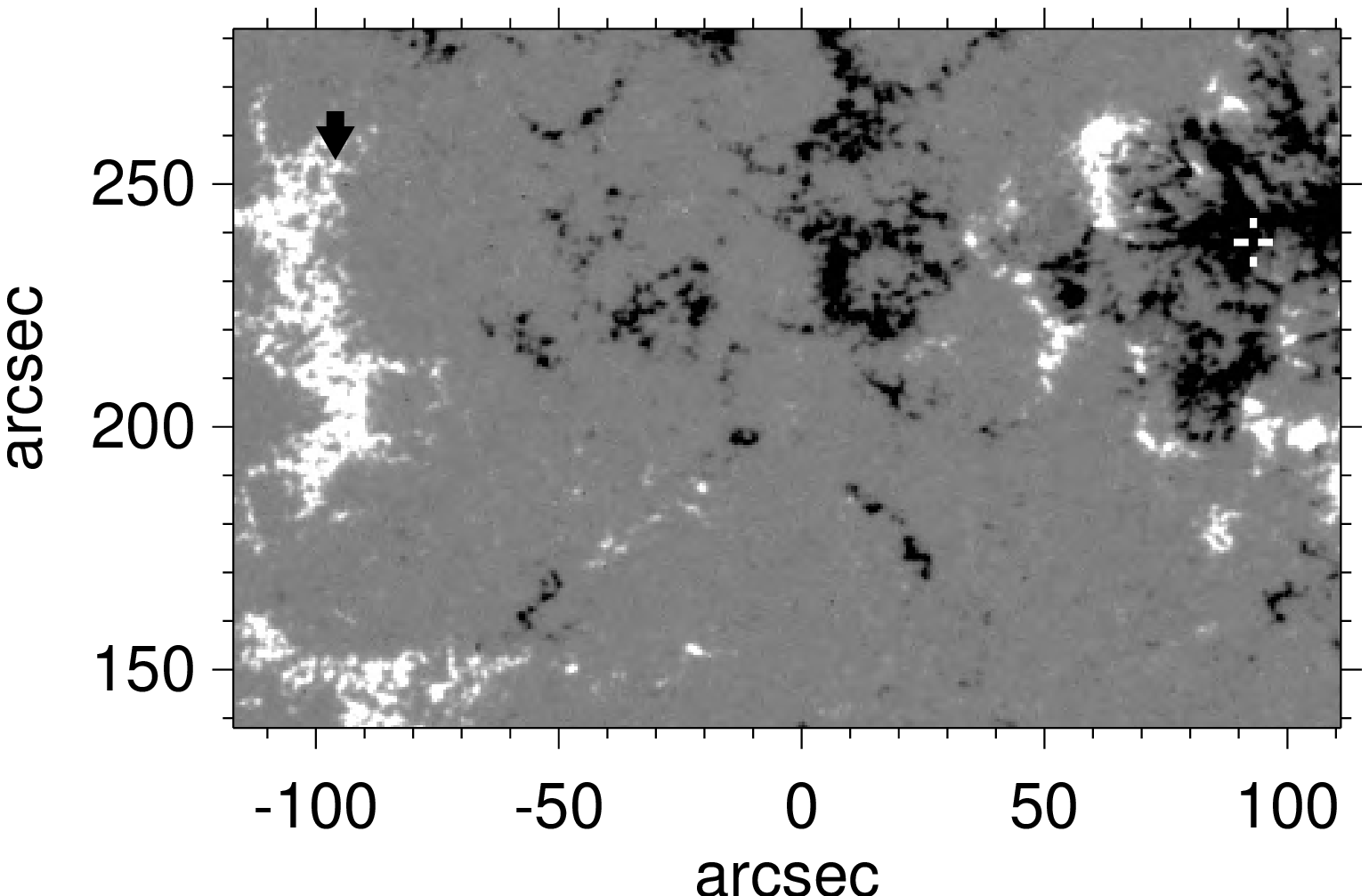}
}
\caption{AIA 171\AA\ (left) and HMI line-of-sight (right) images acquired at 17:15 UT on 10 May 2016. The arrow (-97\arcsec,225\arcsec) and the cross-hairs (92\arcsec,238\arcsec) mark the position of the remote and the sunspot footpoint of a fine active loop.  
\textbf{An animation of the AIA 171\AA\ images is available in the online Journal. The animation runs from 16:50 to 17:30 UT and includes white arrows marking the positions of the remote and the sunspot footpoint of a fine active loop.} 
\label{aia_hmi}}
\end{figure}

\begin{figure}
\centering{  
\includegraphics[width=2.5truein]{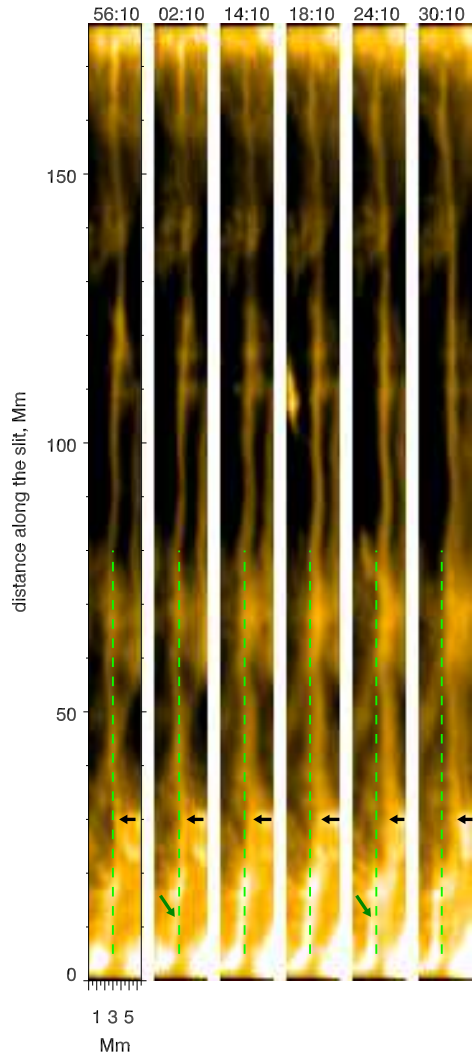}
}
\caption{Evolution of the AIA 171\AA\ loop between 16:56:10~UT and 17:30:10~UT on 10 May 2016. The title above each panel indicates the acquisition time (MM:SS, no hours) of the corresponding image. The lower end of the loop at y=3~Mm is the sunspot footpoint (west), while the remote plage footpoint (east) is at y=175~Mm. The green dotted line segments mark the position of the loop as measured at 16:56:10~UT. The black arrows point to the current position of the loop which was gradually shifting to the right (south). The two green arrows indicate the evolving foot of the loop discussed in the text.}
\label{xt_panels}
\end{figure}

The May 10 loop seen in the left panel of Figure \ref{aia_hmi} and the \textbf{online} movie showed two revival episodes taking place between 16:40~UT and 17:30~UT. The remote loop foot-point was situated in a positive polarity plage (vertical arrow), while the other foot-point was anchored in a sunspot-adjacent area riddled with small pores (cross-hair). The first episode began at 16:40~UT and peaked by approx 16:50~UT, when the entire loop became visible in AIA~171~\AA\ images. The loop intensity soon rapidly decreased only to revive again at 17:05~UT and it became well defined and traceable from one footpoint to another by 17:20~UT. During this 15~min interval the 171~\AA\ sunspot footpoint brightened and the sunspot adjacent half of the loop appeared brighter and extending as well, giving the impression of solar plasma being injected into the loop at that footpoint. In Figure \ref{xt_panels} we show snapshots of a straightened loop as it evolved. We straightened loops by stacking 6\arcsec wide image slices cut at every pixel along the spine of the loop in the direction orthogonal to the local tangential of the loop. In each panel, the green vertical dashed line marks the initial position of a loop leg at the onset of the event, while the arrows indicate the subsequent displacements of that leg. As the loop evolved the displaced leg soon appeared detached by $\sim$ 2~Mm from the bright AIA 171~\AA\ footpoint (shifted away from the vertical line) and instead a bright $\sim$20~Mm jet like feature appeared at that location (panel 24:10). The observed loop displacement and the new jet feature are consistent with an continuous interchange type of reconnection where an open field line may ``jump'' over a dipole that emerges next to it \citep[e.g.,][]{2002JGRA..107.1028C,cranmer_2010,2041-8205-863-2-L22}.

\begin{figure}
\centering{
\includegraphics[width=3.truein]{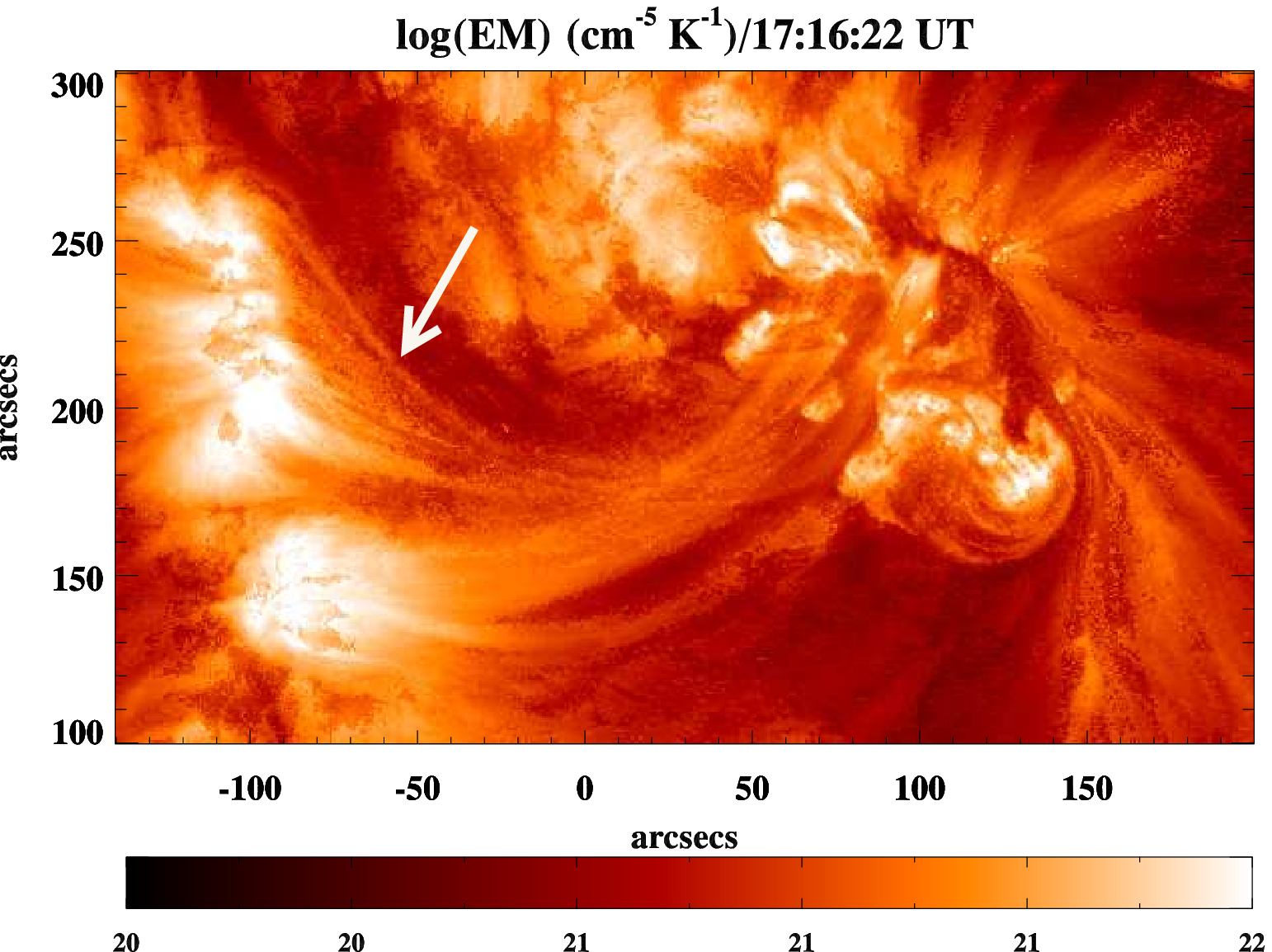}
\includegraphics[width=3.truein]{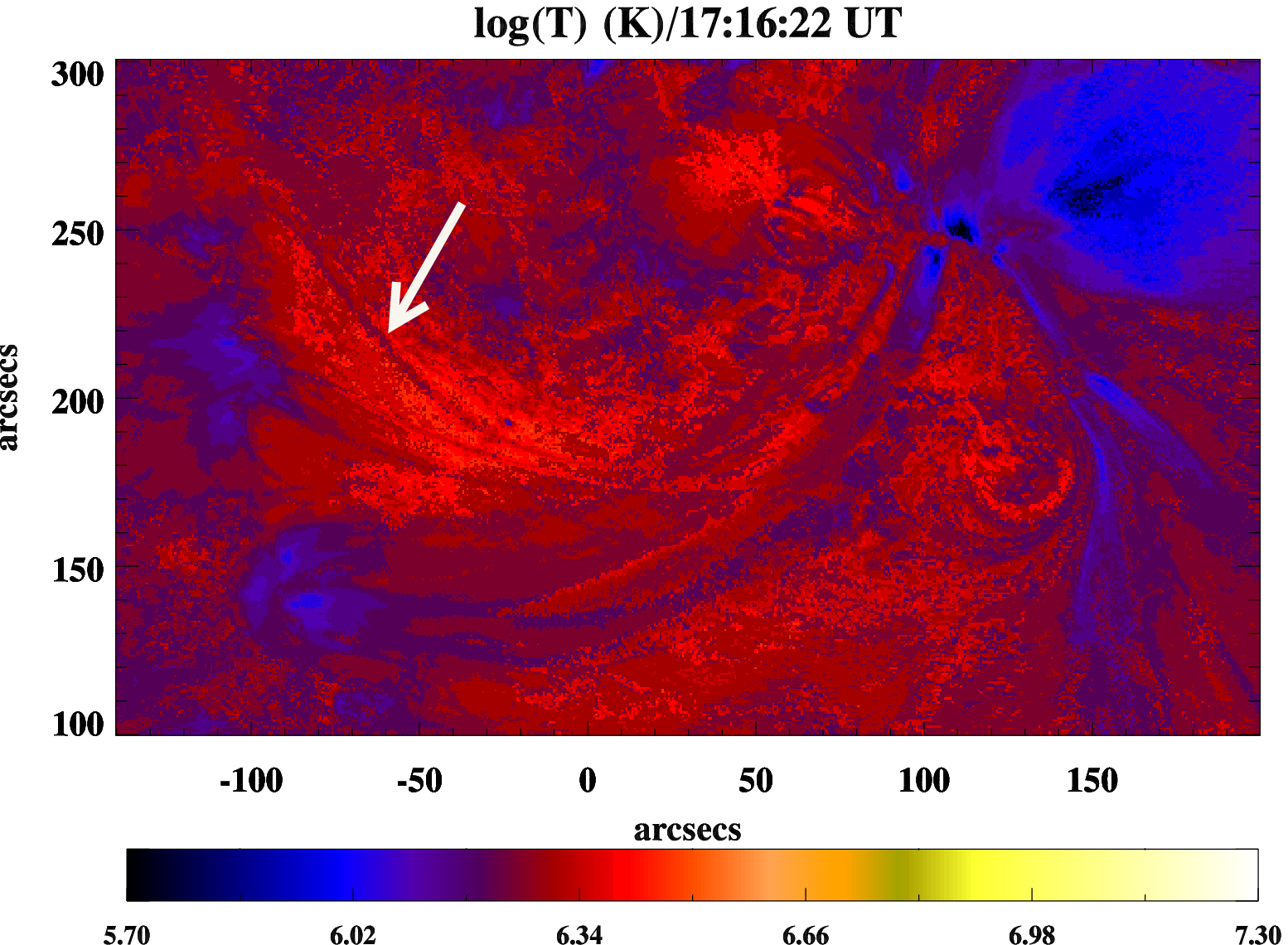}}
\caption{Emission measure, $\log EM$, (left) and $\log T$ (right) determined for Loop I (indicated by the arrow) at 17:16:22~UT.}
\label{aia_temps}
\end{figure}

To determine differential emission measure (DEM) distribution of an AR, we used the automatic DEM code developed by \cite{2013SoPh..283....5A}. The code first co-aligns near simultaneous images from six AIA channels and then calculates the Gaussian DEM distribution with best-fit values for the peak emission measure, peak temperature, and Gaussian temperature widths in each pixel. In Figure \ref{aia_temps} we show emission measure, $\log EM$, and peak temperature, $\log T$[K], maps calculated near the peak of the loop evolution (17:16 UT). Although the thermal image of the right (sunspot) half of the loop is diffuse, the $\log T$ map shows that the loop temperature was nearly uniform in the 1.0-1.5~MK range that put it in the class of warm loop \citep{1999ApJ...517..497L,2003A&A...406L...5D}.

\begin{figure}
\centering
\includegraphics[width=3.5truein]{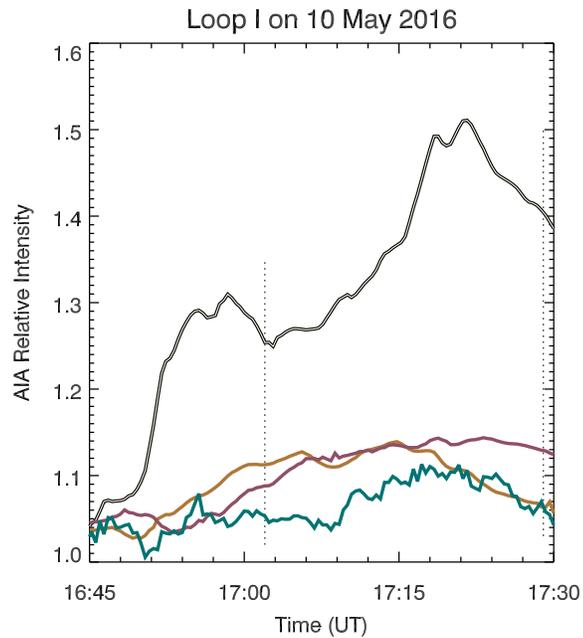}
\caption{AIA 131\AA\ (cyan), 211\AA\ (purple), 193\AA\ (gold), and   171\AA\ (double black) light curves determined near the apex of Loop I (box in Figure \ref{aia_hmi}). Vertical dashed line indicate the studied time interval.}
\label{aia_l1}
\end{figure}

We analyzed AIA 94, 131, 211, 193, 171, 335 \AA\ images to understand heating and cooling processes in the loop and to determine the temporal delay in the peak intensities among different channels \citep{2011ApJ...738...24V}. In Figure \ref{aia_l1} we plot AIA intensity profiles measured near the loop apex within the boxed area shown in Figure \ref{aia_hmi}. {\bf The profiles were normalized relative to the minimal intensity detected during the time interval of interest. To probe intensity variations within the loop we tested several locations and chose the part of the loop that was not affected by the background emission. Moreover, this part was also free from the long lasting foot-point emission, which allowed us to detect subtle emission variations associated with propagation of plasma along the loop.} The AIA 131~\AA\ channel (cyan) shows a nearly steady profile until about 17:09~UT when a small enhancement occurred that peaked at approximately 17:18~UT and consisted of numerous weak impulsive events. The AIA 171~\AA\ profile (double black) generally agrees with the AIA 131~\AA\ since it represents contributions from both cool and hot temperature plasma. The AIA 211~\AA\ profile (purple line) is also similar to the above two profiles in that it peaks near 17:18~UT although the fine scale structure of the profile is different. To the contrary, the 193~\AA\ profile (gold) behaves quite differently exhibiting an decreasing trend during the peak times 17:18-17:24~UT) in the hotter channels. We also note that the plasma flow pattern observed in the AIA 211, 193, and 131~\AA\  was quite similar to that seen in the 171~\AA\ channel (i.e., from one footpoint to the other). The loop and plasma flows were not detected in the hot AIA 94 and 335~\AA\ channels.

The time lag in various UV light curves and its magnitude are important for understanding the physical processes behind the loop heating and cooling episodes. In case of a single heating event, the heated plasma is first expected to be detected in hotter UV spectral lines such as 131\AA\ or 211\AA\ and, as the plasma cools, the loop may become visible in progressively cooler lines such as 171\AA\ \citep[see e.g.,][]{2011ApJ...738...24V}. When several consecutive heating events occur at a time interval that is much shorter than the plasma cooling time, then the time lag between the cool and hot channels is less pronounced (if at all present) since the loop may be reheated again as it cools. In this case the heating is considered to be steady and the emission is thought to be hot so that the cooling sequence mentioned above may not be observed. The studied event does not show a well defined time lag and appears to be relevant to the idea of a steady heating process.

\begin{figure}
\centering
\includegraphics[width=6.5truein]{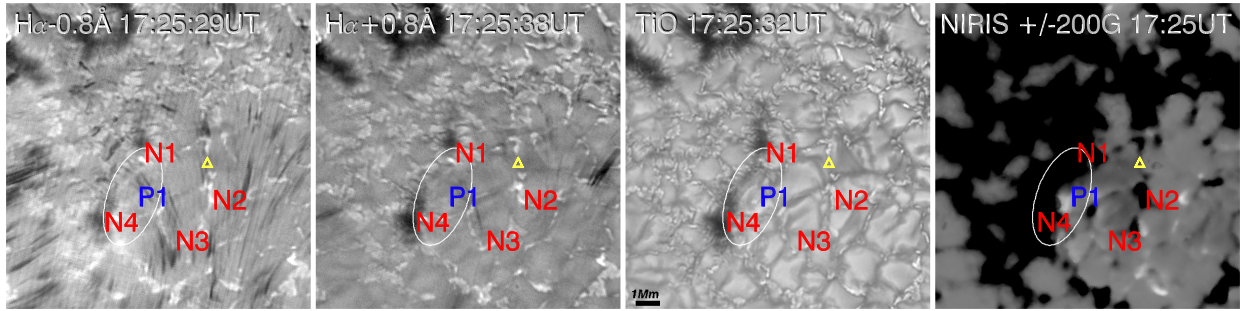}
\caption{Chromospheric VIS H$\alpha-0.8$\AA\ and H$\alpha+0.8$\AA\ images (two left panels), photospheric TiO 7059\AA\ image, and NIRIS B line-of-sight component of the magnetic field (right) in the vicinity of the AIA 171~\AA\  sunspot loop footpoint, location of which is outlined by the ellipse. ``N1'' and ``N4'' mark two negative polarity magnetic elements and the co-spatial compact brightenings, while ``P1'' marks a small positive polarity flux that situated in the middle of a granule. ``N2'', ``N3'', and ``$\triangle$'' are plotted here to ease image comparison. The FOV of each panels is 17$\arcsec$.4$\times$17$\arcsec$.4~Mm.}
\label{hationiris}
\end{figure}

\begin{figure*}
\centering{
\includegraphics[width=5.3cm]{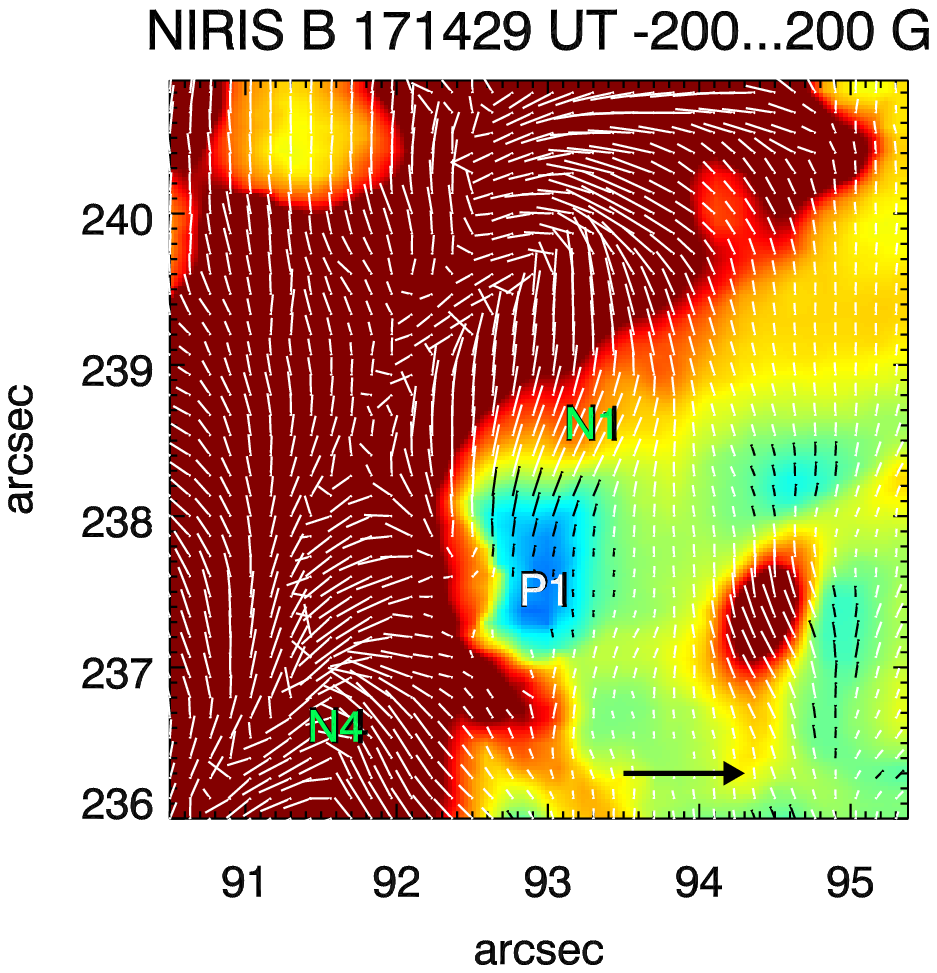}
\includegraphics[width=5.3cm]{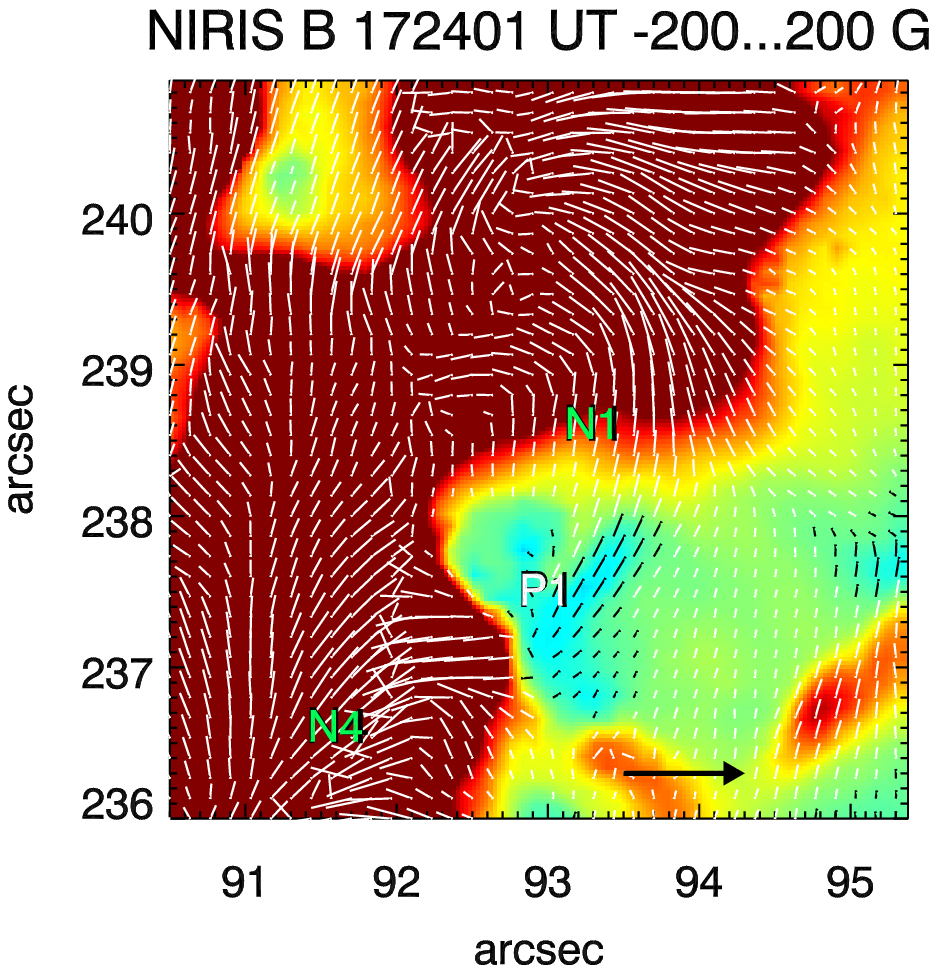}
\includegraphics[width=5.3cm]{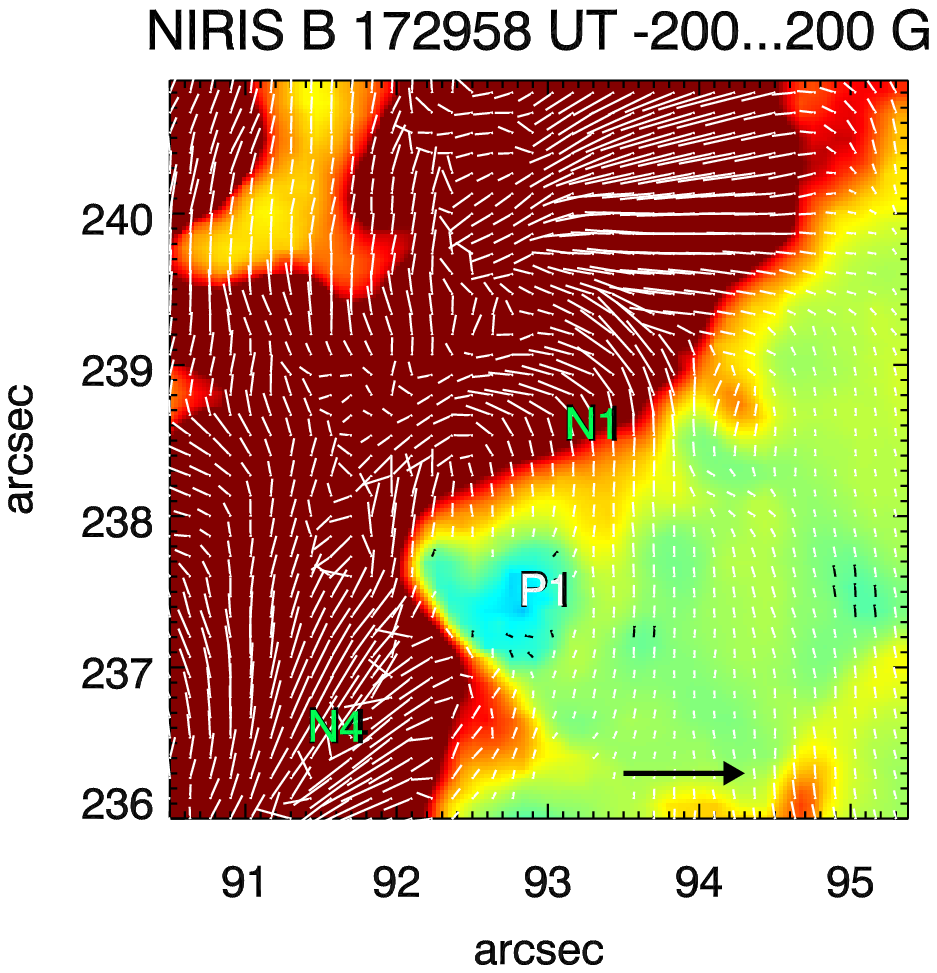}
}
\caption{Evolution of the magnetic environment at the sunspot footpoint of the loop. The black arrow represents 2000 G transverse fields. The LOS fields (background) are scaled between -200 G (red) and 200 G (blue). ``N1'', ``P1'', and ``N4'' mark position of magnetic elements under discussion. The extended AIA 171~\AA\ sunspot loop footpoint occupied space between N1 and N4 (also see Fig. \ref{hationiris}).}
\label{vmf}
\end{figure*}

In Figure \ref{hationiris} we show the photosphere and lower chromosphere associated with the loop sunspot footpoint. The footpoint position, seen as a bright AIA 171~\AA\ patch, is outlined in these panels by the ellipse. Co-alignment of GST and SDO data showed that it was associated with negative polarity elements N1 and N4 as well as a positive polarity fields P1. {Note that the NIRIS observations began at 17:14 UT. According to TiO images, the P1 element was moving toward N4 and it was part of a new flux emergence that began at least at 17:09~UT.} It also created a footprint in photospheric TiO images seen as a faint dimmer ``indentation'' atop of a granule with a brighter round feature in the middle (seen on the left of the letter ``P''). Co-temporal GST/VIS images did not show any distinct jetting that could be reliably linked to the flux emergence event besides possibly faint narrow blue-shifted jets seen between ``P1'' and ``N1'' in the GST/VIS \ha-0.8\AA\ image. In Figure \ref{vmf} we show evolution of the GST/NIRIS vector magnetic fields associated with the sunspot footpoint. According to the data, N1 and P1 were magnetically connected with the transverse fields running along the P1-N1 line. As the flux emergence progressed, P1 shifted toward N4 and the N1 flux increased. However, by 17:30~UT the transverse fields connecting the N1-P1 pair seem to have weakened, while the P1-N4 connection has strengthened as evidenced by the transverse fields that now became oriented along the line connecting the elements. This orientation change as well as weakening of P1 both suggest that we witnessed a flux cancellation process driven by emergence of small-scale fields at the borderline of mostly unipolar plage fields.

\begin{figure}
\centering{
\includegraphics[width=15cm]{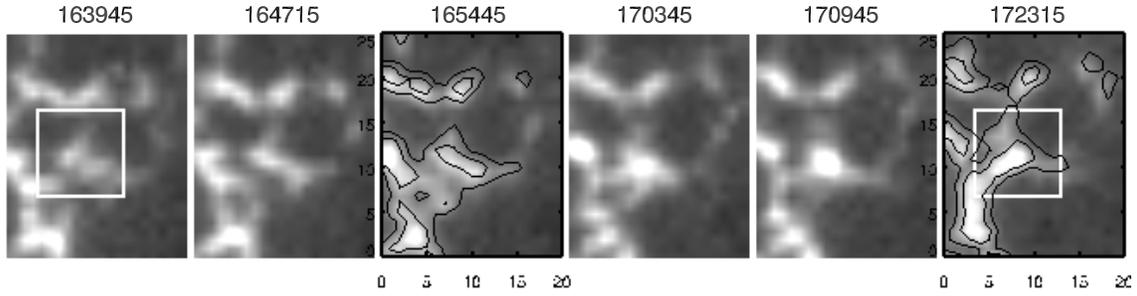}}
\caption{Evolution of HMI.M\_45s series magnetic fields at the remote footpoint of Loop I. The footpoint location is enclosed by the box. Only the positive (white) fields inside the box, showed considerable variations between 16:39~UT and 17:23~UT, while the rest of the features remained in a nearly stable state.}
\label{hmi_remote}
\end{figure}

\begin{figure}
\centering{
\includegraphics[width=4.4truein]{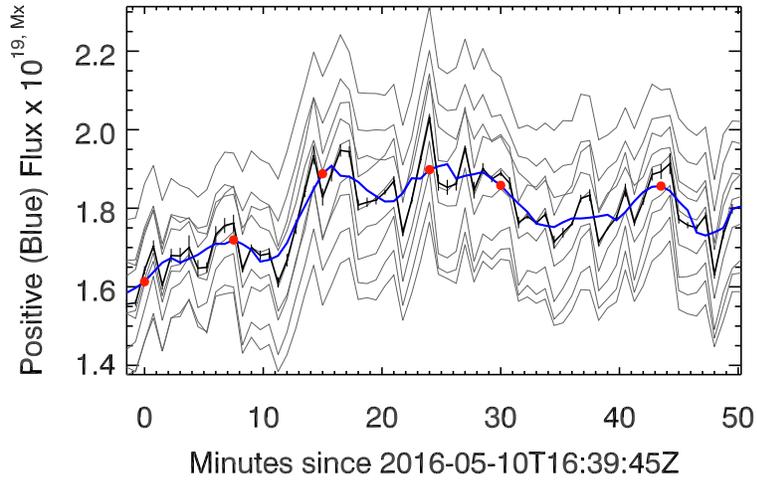}}
\caption{Time variations of the positive flux calculated over the entire FOV shown in Fig. \ref{hmi_remote}. Thin gray lines show flux calculated separately for 9 positions of the bounding box. The thick black line show the corresponding averaged profile, while the blue line is the smoothed averaged profile. For error bars calculations see text. The red closed circles on the blue curve indicate times of the panels in Fig. \ref{hmi_remote}.}
\label{hmi_remote_flux}
\end{figure}

Figure \ref{hmi_remote} shows evolution of HMI LOS fields at the remote footpoint of Loop I (arrow in Figure \ref{aia_hmi}). Comparison of 16:39:45, 17:03:45 and 17:23:15~UT panels suggests that a positive polarity field (enclosed by the white box) has notably increased during that period. We should note that during the analyzed time interval the remote footpoint did not show any detectable AIA 171~\AA\ brightening variations. Considering that other magnetic elements remained mostly unchanged and there are no signs of magnetic elements being displaced by convective plasma flows, we are inclined to interpret the positive polarity increase as a result of flux emergence rather than field enhancement due to converging flows. The blue line in Figure \ref{hmi_remote_flux} is a 5-point smoothed version of the averaged positive LOS flux (thick black line) calculated over the entire FOV shown in Figure \ref{hmi_remote} and it is evident that beginning 16:40~UT the flux experienced a rapid 12\% increase. The negative flux in this area did not show any regular trend and was negligible to be shown in the plot. In order to estimate errors we first determined flux time profiles for nine different positions of the bounding box, which was sequentially shifted by one pixel in various directions. Their average profile is plotted in Figure \ref{hmi_remote_flux}. The resulting profiles (gray thin lines) had a very similar structure, however their mean values were quite different due to the fact that some flux was entering or leaving the box as it shifted. While this edge flux did not affect the time variations, variations of the mean were too large to directly estimate the errors, so we subtracted the respective mean from each of nine profiles and then calculated the r.m.s. using the residual (detrended) profiles. Thus, the small error bars indicate the stability of these individual time profiles rather than the absolute scatter of the data points induced by the edge effect.

In order to further argue for the flux emergence idea we refer to the fact that nearly always one polarity in the emerging flux is compact and much stronger than the other one \cite[e.g.,][] {1981phss.conf....7M,1998A&A...333.1053L,0004-637X-601-1-530,nst_jet_2010,2015LRSP...12....1V,2017ApJ...846..149C}. In this case the emerging minor (relative to the positive polarity plage) negative polarity may have been scattered and thus largely below the detection limit of the HMI instrument. Additionally, it could have partially cancelled with the surrounding positive polarity fields as it was emerging, which would further contributed to the deficit of the negative flux.

{To summarize, the analyzed data showed notable magnetic activity at both footpoints of this loop consistent with the flux emergence, though solid observational evidence of flux cancellation exist only for the brighter sunspot footpoint. If the loop indeed was filled with plasma evaporated by reconnection driven by the new flux emergence at the sunspot end of the loop that began shortly before 17:09~UT, }then we estimate that the density enhancement propagated along the loop at a rate of $\sim$ 110~km~s$^{-1}$, accepting that loop brightness at y=90~Mm (loop midpoint, Figure \ref{xt_panels}) enhanced at approx 17:18~UT. This estimate is somewhat higher than the 40-60~km~s$^{-1}$ rate reported from spectroscopic observations \cite[][]{2002ApJ...567L..89W,2009ApJ...694.1256T,2015ApJ...805..167S} and is consistent with \cite{2008A&A...482L...9O} measurements.

\subsection{17 Dec 2015 Loop}

\begin{figure}
\centering{
\includegraphics[width=3.2truein]{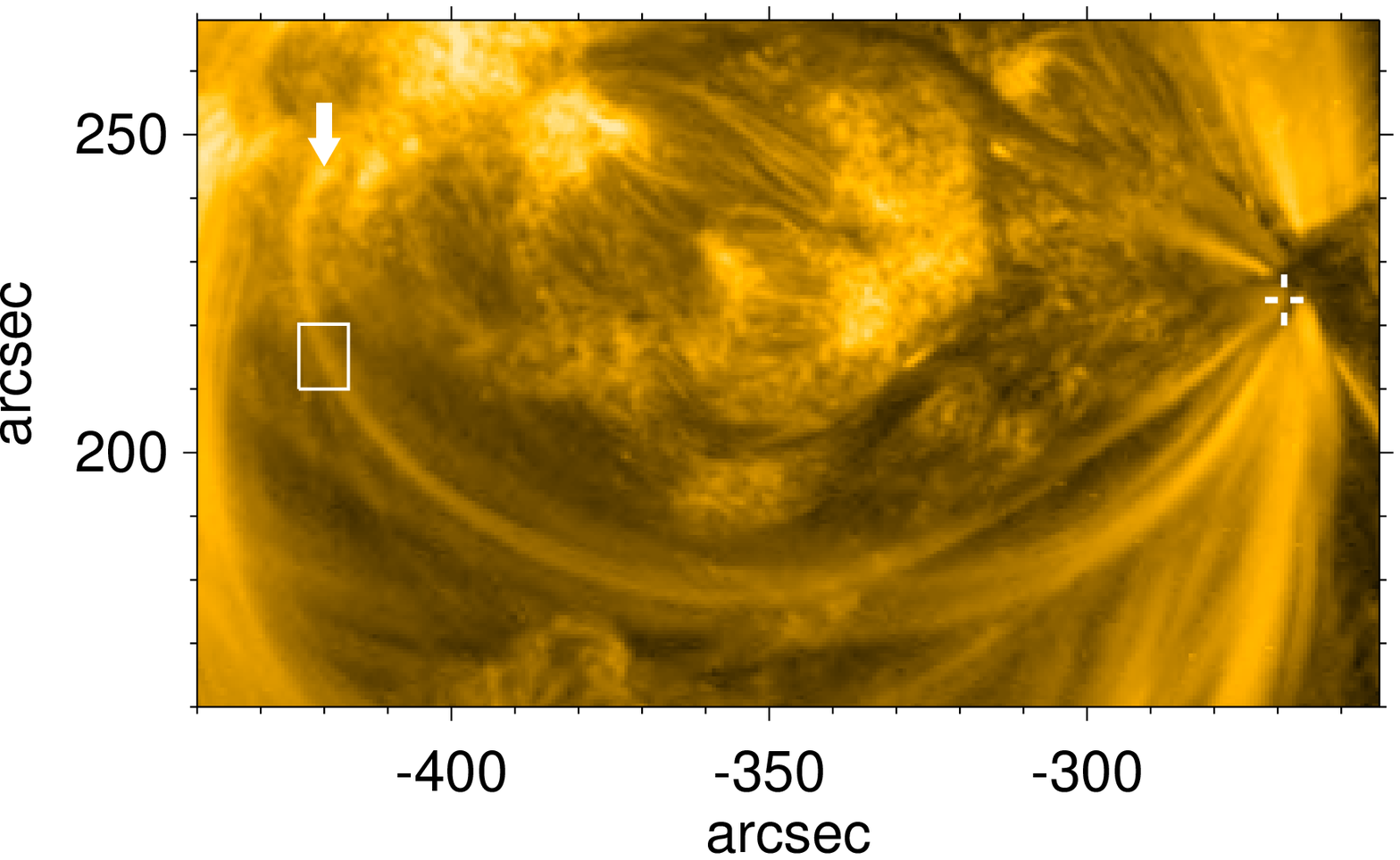}
\includegraphics[width=3.2truein]{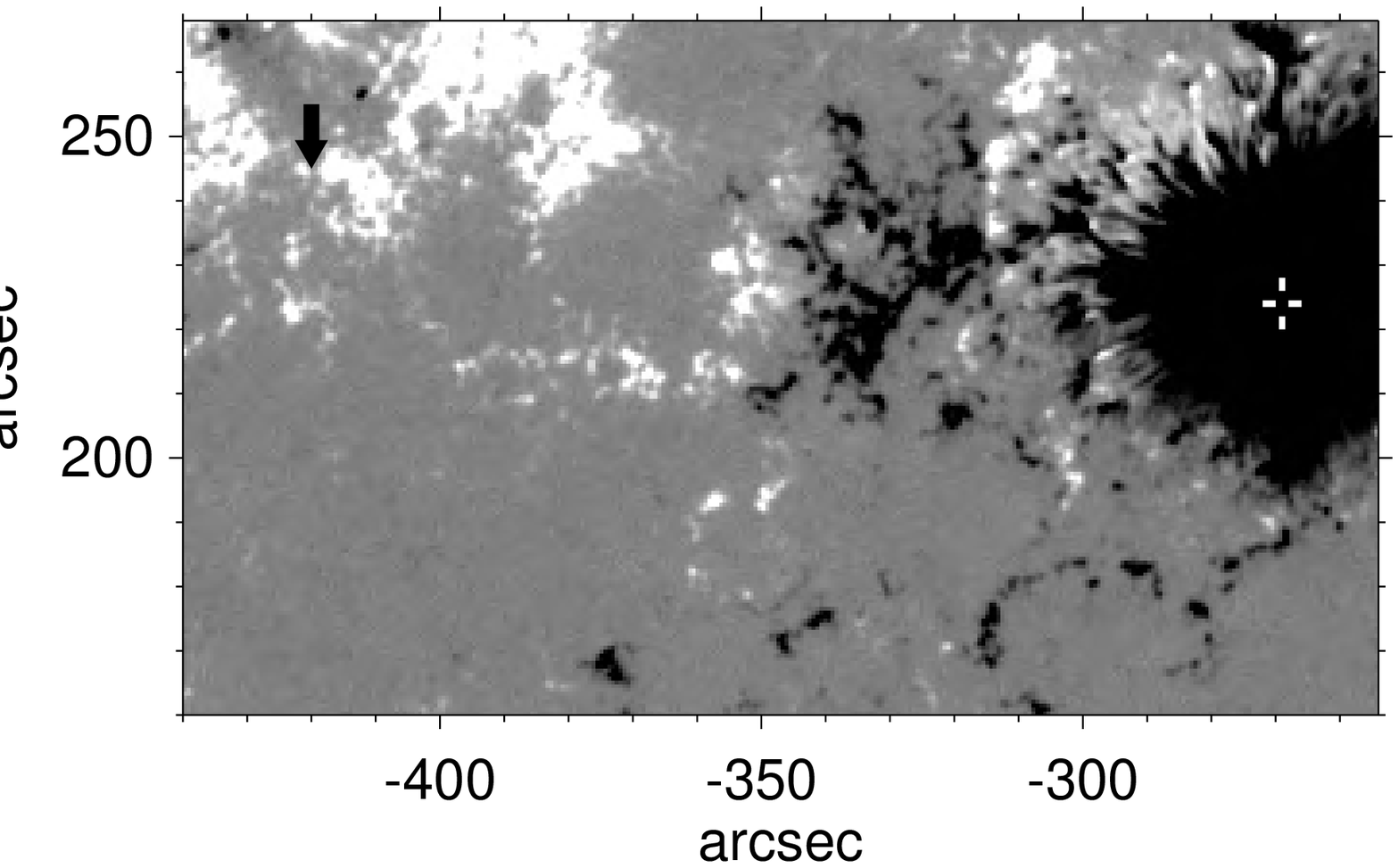}}
\caption{AIA 171\AA\ (left) and HMI line-of-sight (right) images acquired at 17:30 UT on 17 Dec 2015. The arrow (-420\arcsec,245\arcsec) and the cross-hairs (-270\arcsec,225\arcsec) mark the position of the remote and the sunspot footpoint of a fine active loop. 
\textbf{An animation of the AIA 171\AA\ images is available in the online Journal. The animation runs from 17:00 to 17:40 UT and includes white arrows marking the positions of the remote and the sunspot footpoint of a fine active loop.} 
\label{aia_hmi2}}
\end{figure}

\begin{figure}
\centering
\includegraphics[width=3.5truein]{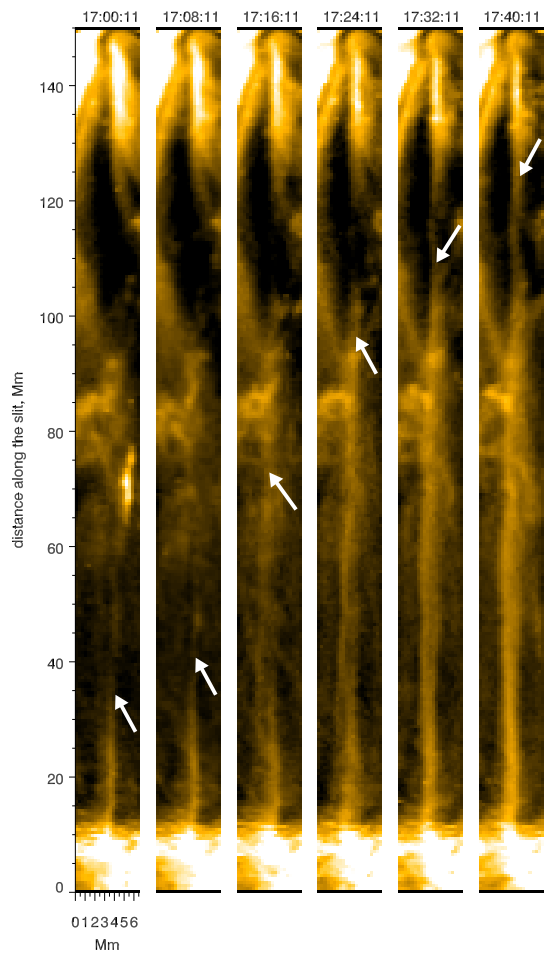}
\caption{Evolution of the AIA 171\AA\ loop (arrows) between 17:00~UT and 17:40:10~UT on 17 Dec 2015. The panel title indicates the acquisition time in UT of the corresponding image. The lower end of the loop at y=7~Mm is the remote footpoint (east), while the sunspot footpoint (west) is at y=147~Mm. The arrows indicate the position of the plasma enhancement front.}
\label{xt_panels_loop2}
\end{figure}

Loop II was connecting a plage region with an outer umbral region of the main leading sunspot in NOAA AR 12470 (Figure \ref{aia_hmi2} and \textbf{online} movie), which was an ALMA campaign target \citep{2017ApJ...841L...5S}. Although GST observations did cover the leading sunspot we do not use them here because of unsatisfactory seeing quality. Loop II is similar to Loop I in size and life time, however, one notable and intended difference is that its sunspot footpoint was rooted in the umbra as opposed to a near-sunspot area as in case of Loop I. Also, while Loop I was apparently filled with plasma nearly simultaneously along the entire length (possibly because plasma injection occurred at both footpoints), Loop II showed propagation of a dense plasma front from the remote footpoint toward the sunspot (Figure \ref{xt_panels_loop2}, arrows). This flow pattern is consistent with the idea of energy release occurring at one of the footpoints of the loop which injects hot plasma into the loop and it further travels the other footpoint of the loop. It has been reported that such energy release events at loop footpoint not only lead to rapid heating but also trigger a longitudinal compressive wave along the loop, which may bounce back and forth several times before fading \citep{2013ApJ...779L...7K, 2015ApJ...804....4K}.

\begin{figure}
\centering
\includegraphics[width=3.truein]{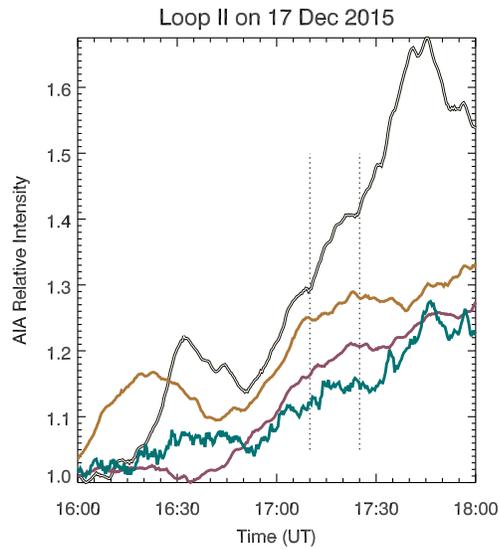}
\caption{AIA 131\AA\ (cyan), 211\AA\ (purple), 193\AA\ (gold), and   171\AA\ (brown) light curves determined near the remote footpoint of Loop II. Vertical dashed line indicate the studied time interval.}
\label{aia_l2}
\end{figure}

Figure \ref{aia_l2} shows AIA intensity profiles determined close to the endpoint of the loop (see boxed area in Figure \ref{aia_hmi2}). The loop has experienced several revival episodes as well. The first episode occurred after 16:00~UT with the peak 211~\AA\ intensity (purple) at 16:20~UT. The 193~\AA and 171~\AA\ profiles peaked at 16:20 and 16:30~UT, correspondingly, resulting in about a 10~min time lag, which is consistent with an impulsive heating event. Shortly after that all AIA profiles showed a steady increase with small-scale details suggesting multiple energy injection events. The time interval considered here is marked with two vertical dotted lines. {According to the figure, by 17:10~UT the 193~\AA\ the intensity (gold) have reached a turning point at which it slightly decreased and then continued to grow but already at a different rate (compare 16:55-17:10 and 17:10-17:20~UT intervals) until 17:20~UT, followed by intensity decrease till 17:40~UT. At the same time, the 171~\AA\ intensity gradually increased and it peaked at 17:45~UT.} Note that during this time interval all curves showed a steady growth without any pronounced time lag in their profiles.

\begin{figure}
\centering
\includegraphics[width=6.5truein]{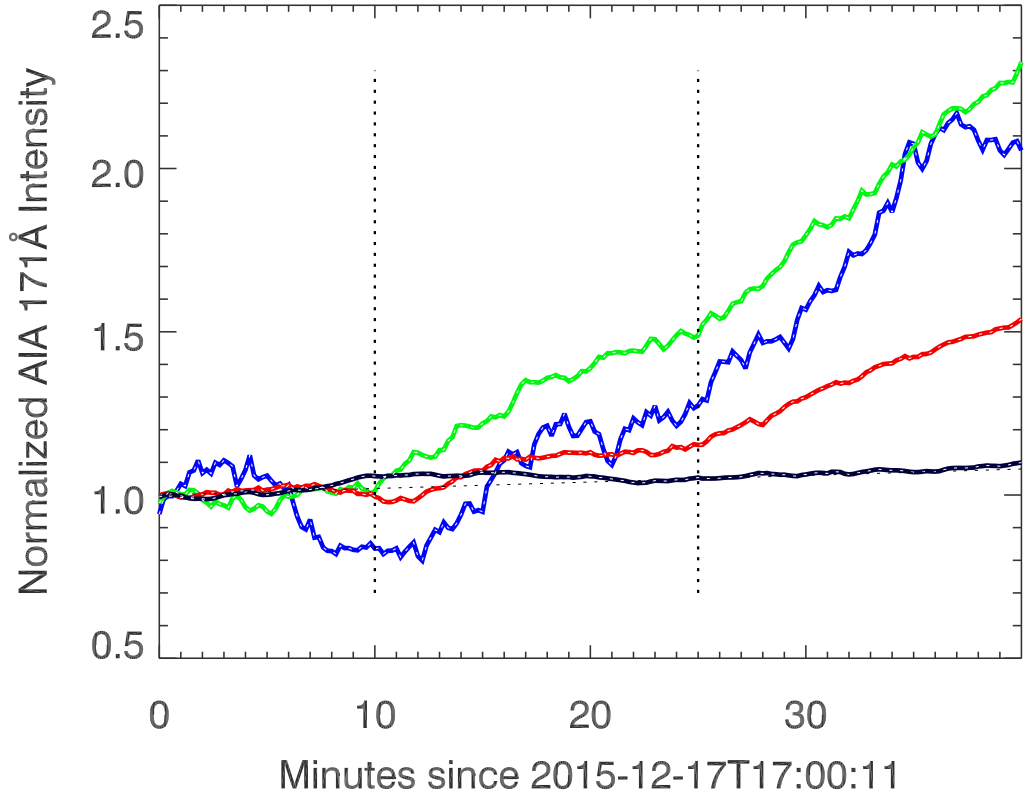}
\caption{Normalized intensity profiles as measured the loop footpoint at positions y=10 (black line), y=25 (red), y=40 (green), and y=110~Mm (blue, see Fig. \ref{xt_panels_loop2}). The vertical dotted lines indicate the start time of density enhancement and the time when the dense plasma front reached y=110~Mm mark in Figure \ref{xt_panels_loop2}.}
\label{xt_profiles}
\end{figure}

\begin{figure}
\centering{
\includegraphics[width=4.6truein]{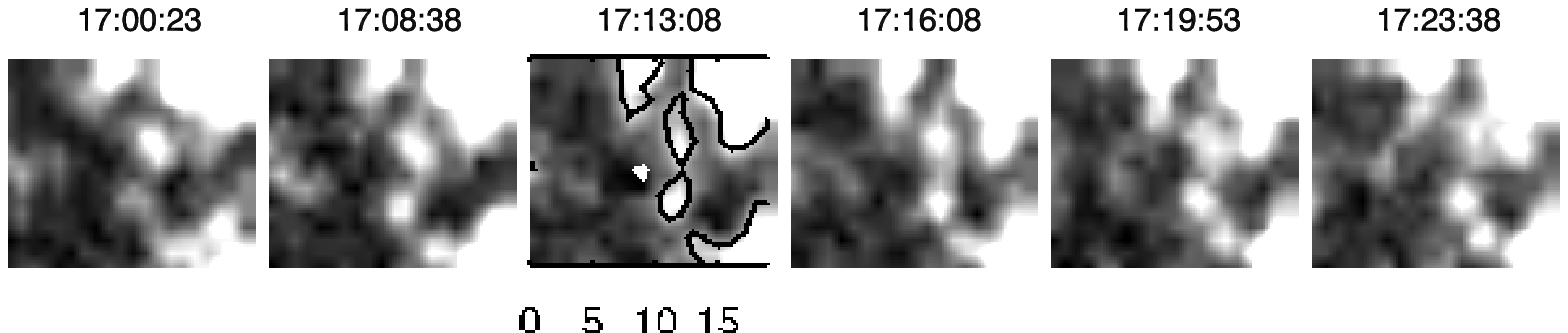}
\includegraphics[width=1.7truein]{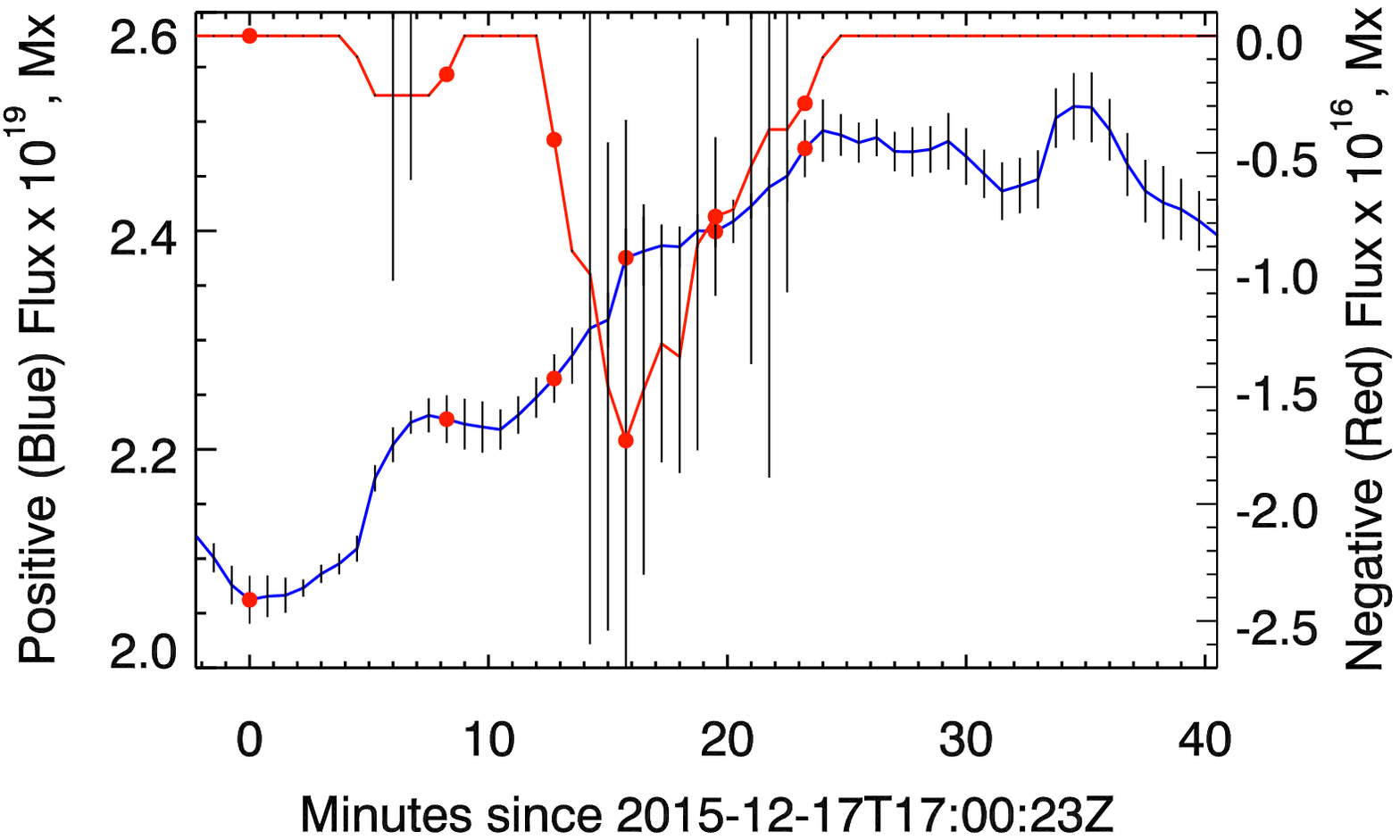}}
\caption{Evolution of the photospheric HMI.M\_45s series magnetic fields at the remote footpoint of Loop II. Black contours are drawn at 100 and 500~G levels, while the white contours are plotted at the 5~G level. The units shown on the x-axis in the third panel are HMI pixels. The field of view is 19$\times$16 pixels, which corresponds to $6.9\times5.8$~Mm. The right panel shows time profiles of the positive (blue) and negative (red) flux calculated over the entire FOV. The red closed circles on the time profiles indicate times of the panels.}
\label{hmi_remote2}
\end{figure}

Figure \ref{xt_profiles} displays loop intensity profiles at four positions along the loop. The intensity profile measured at the loop footpoint (black) shows an increase at 17:07~UT, which coincides with the peak of the negative HMI flux (Figure \ref{hmi_remote2}) and suggests that flux cancellation began about 10~min after the onset of the emergence process. The footpoint intensity returned to the pre-event level at 17:22~UT, the time when the negative flux completely diminished and the injected plasma reached the sunspot footpoint. At the same time all AIA profiles began to decline as well (Figure \ref{aia_l2}). The loop intensity at the sunspot footpoint peaked at 17:37~UT after which it began to decrease. The 171\AA\ emitting plasma began to appear at about 17:10~UT at the remote footpoint at y=40~Mm (Figure \ref{xt_panels_loop2}) and reached position of y=110~Mm along the loop by $\sim$ 17:25~UT, which resulted in a 80~km~s$^{-1}$ rate, which is in agreement with that derived for Loop I. The green and blue curves show intensity variations at y=40~Mm and y=110~Mm, correspondingly. Although intensity at y=110~Mm was gradually increasing (blue curve), only at t=25~min it became possible to clearly identify the loop. 

\begin{figure}
\centering
\includegraphics[width=3.truein]{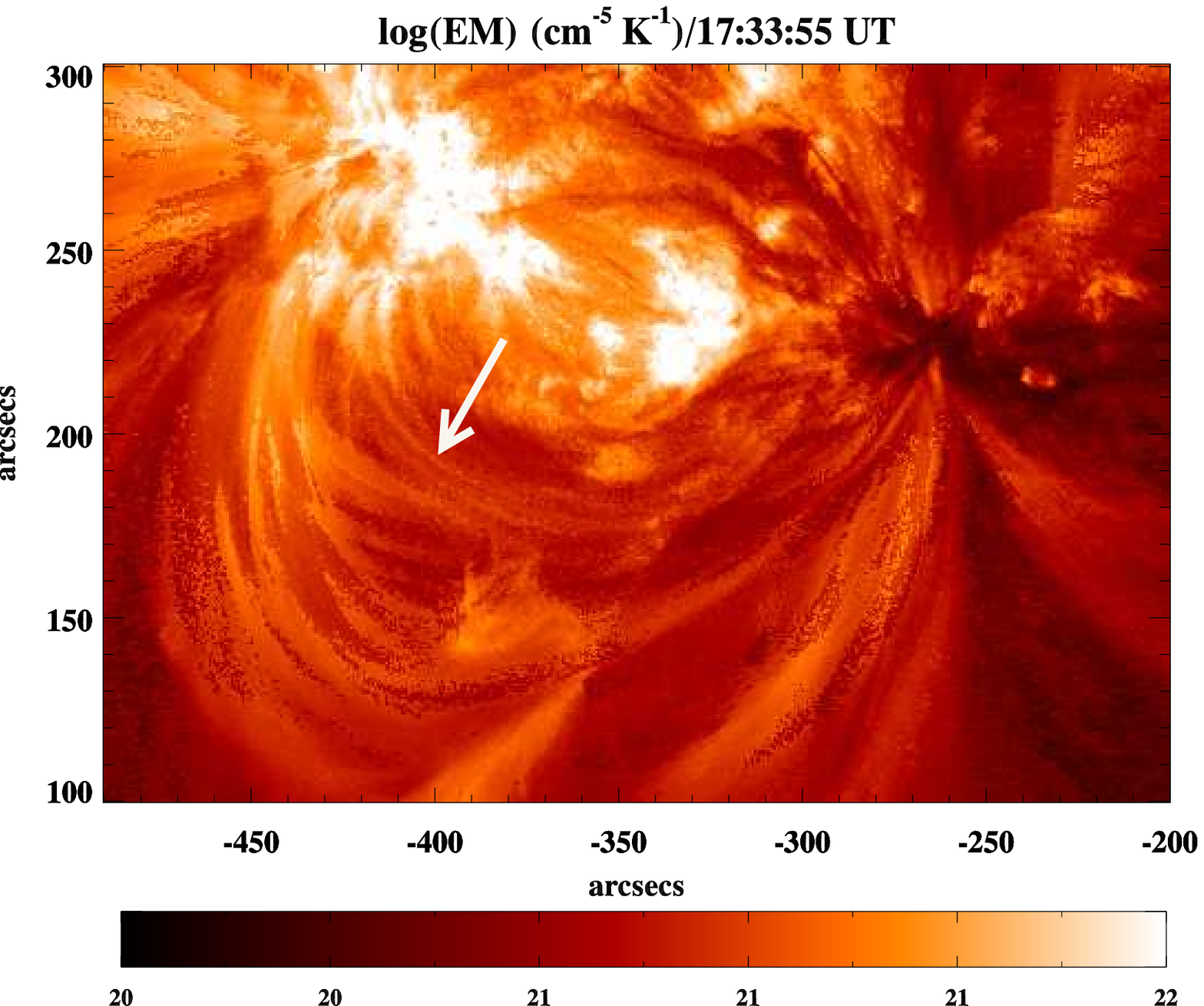}
\includegraphics[width=3.truein]{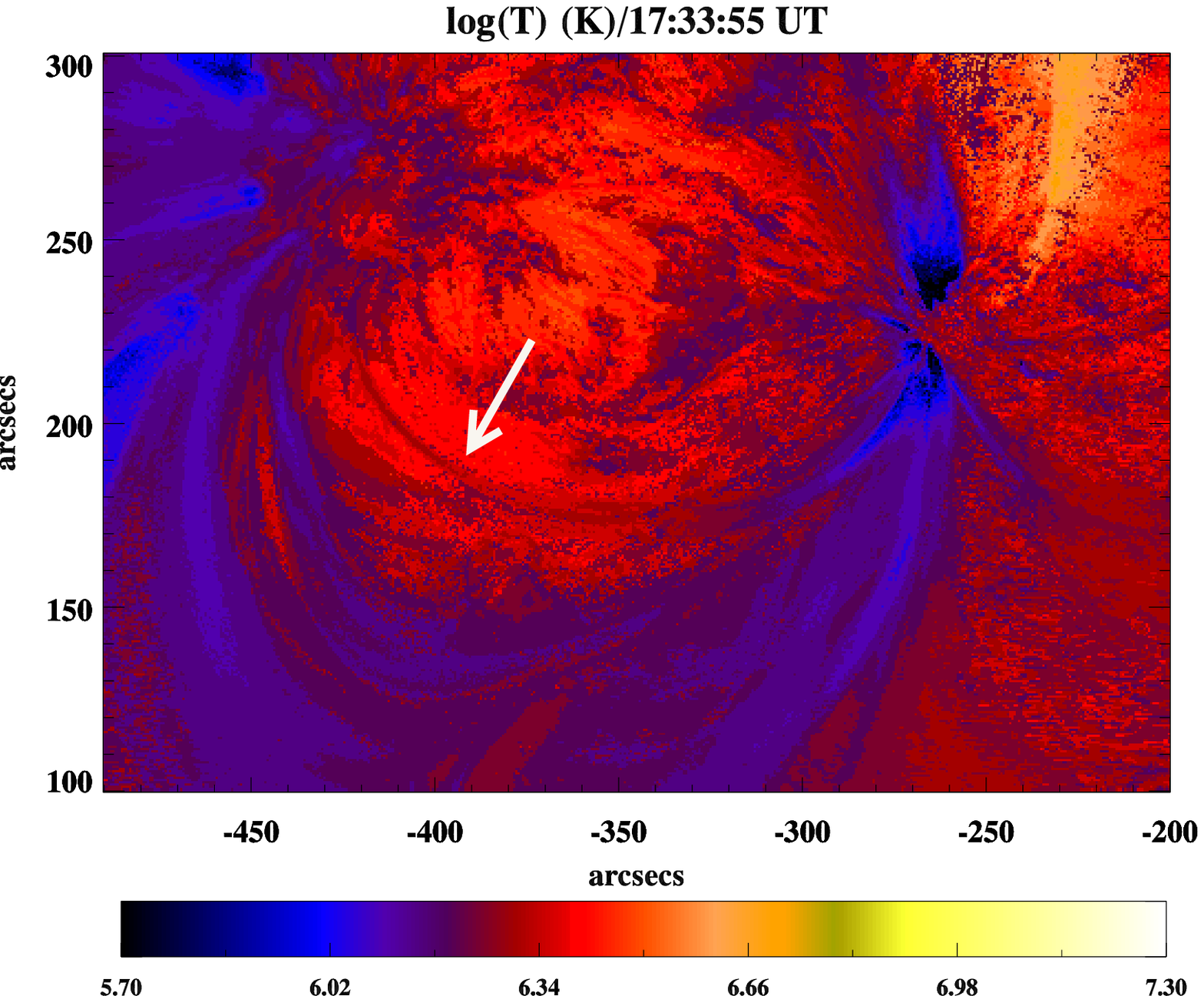}
\caption{Emission measure (left) and $\log(T)$ (right) determined at 17:33~UT for Loop II. The loop top appears to be somewhat hotter reaching 1.5~MK (arrow) while the rest of the loop was found to be at temperatures slightly below 1.0~MK.}
\label{aia_temps2}
\end{figure}

In Figure \ref{aia_temps2} we show $\log EM$ and $\log T$[K] maps calculated for Loop II at 17:33~UT. Unlike the Loop I case, only ``remote'' half of the loop exhibited enhanced temperatures of $(1.0-1.5~MK)$ and $\log EM$, which too may be classified as a warm loop \citep{1999ApJ...517..497L,2003A&A...406L...5D}. The sunspot footpoint was located in a large sunspot with its umbra partially fragmented by several thin LBs, however, it was not anchored at an LB but in an outer uniform umbral area. Although seeing quality of GST data collected at that time was not very good, neither these data nor AIA images showed any signatures of jetting at that umbral location. Available \textit{IRIS} data also did not indicate any activity at that footpoint as well. Note, that \textit{IRIS} began to observe this sunspot at 17:33~UT as a part of ALMA campaign. Moreover, we could not find traces of the loop in any of the available \textit{IRIS} data, which makes this case quite different from the one described in \cite{2016A&A...587A..20C}. This led us to conclude that the remote footpoint was responsible for the observed revival of the loop. 

Evolution of HMI LOS fields at the remote loop footpoint is shown in Figure \ref{hmi_remote2}. The loop was rooted at the center of the FOV where a small positive polarity element, outlined by a black contour in the 17:13:08~UT panel, gradually diminished. The graphs on the right quantify the field evolution by showing that the positive flux (blue) increased by about 20\% (0.4$\times$10$^{19}$~Mx) during a 20 min time interval. The HMI data also showed very weak signatures of opposite polarity flux (small white contour in center of the 17:13:08~UT panel) appearing next to the positive polarity element. The negative flux appeared at $\sim$17:10~UT and peaked at 1.5$\times$10$^{16}$~Mx level at 17:15~UT, which nearly coincides with the onset of plasma injection (Figure \ref{xt_panels_loop2}). Shortly after that the negative flux disappeared from the FOV. Although very little details on magnetic fields evolution are available to us, the HMI data nevertheless do suggest that flux emergence and cancellation took place at the footpoint of the loop thus possibly causing the plasma injection.

\section{Summary and Discussion}

In this study we focused on the magnetic environment at four footpoints of two coronal loops and concluded that magnetic flux emergence and cancellation have driven plasma and energy injection into the loops. Loop I was connecting a remote plage area with the core of an AR populated with numerous pores. It appeared to be filled with plasma uniformly and simultaneously along its entire length. GST NIRIS and HMI data showed that there was non-negligible magnetic activity at both their footpoints located at the boundary of seemingly unipolar vast of magnetic fields, where magnetic dynamics is expected to be high. High-resolution NIRIS vector magnetograms showed emergence of a small ($\sim$1~Mm) bipole with strong transverse fields, while HMI data only registered an enhancement of the dominant positive polarity. Loop II was spanning nearly entire AR connecting a plage area with the sunspot umbra. The plasma was injected at the remote plage footpoint and this footpoint also showed considerable variations in the HMI magnetograms. In particular, HMI observations registered appearance of opposite polarity fields several minutes prior to the plasma flow onset. The sunspot footpoint did not exhibit neither magnetic nor plasma activity. We thus suggest that the loop activation, plasma flows and heating processes were driven by magnetic reconnection between a small-scale emerging flux and large-scale fields. In this type of configuration the large scale field line ``jumps'' across a dipole, which may correspond to the the loop displacement clearly observed in the AIA~171~\AA\ data. We also conclude that the sunspot rooted bright loops may be caused not only by LB and umbral dots activity \citep[e.g.,][]{2017ApJ...835..240S,0004-637X-854-2-92} but also by the photospheric dynamics at the remote (non-sunspot) footpoint. 

In spite of the fact that there is a substantial body of literature devoted to coronal loop studies, very few publications were addressing the magnetic structure of loop footpoints. Until very recently magnetic field measurements were dominated by SOHO/MDI, SDO/HMI, and Hinode/SOT instruments and several studies based on these data \citep[e.g.,][]{2003A&A...406L...5D,2009ASPC..415..241U} reported that coronal loops are often connected to highly dynamic but nevertheless unipolar plage fields. However, recently \cite{2041-8205-820-1-L13} argued that HMI instrument does not resolve many small-scale structures so, that mixed polarity may be present below the resolution limit. Also, \cite{ch_parameters} showed that at spatial scales below 2~Mm are highly intermittent and burst-like energy release events are possible. \cite{0004-637X-720-2-1380} concluded that changes in the underlying unipolar magnetic fields could account for heating of warm loops but they are not strong enough to provide energy needed for hot loops. \cite{2012ApJ...750L..25J, 2013ApJ...769L..33Z} and \cite{1674-4527-17-3-25} used 10830~\AA\ data to conclude that energy for heating the upper solar atmosphere comes from inter-granular lanes. These findings are well aligned with \cite{2041-8205-723-2-L185} who found that one of the footpoints of quiet Sun coronal loops is often found inside the dynamic inter-network magnetic fields. Using a similar approach \cite{2017ApJ...843L..20T} also argued that coronal heating may be fueled by vigorous magneto-convection which can braid magnetic field lines and that the heating rate is directly dependent on the field strength in the loop. However, strong fields, such as those found in the sunspot umbra suppress magnetoconvection thus reducing the heating rate \citep[e.g.,][]{2014A&A...564A..12C,2017ApJ...843L..20T}. \cite{2017ApJS..229....4C, 2018A&A...615L...9C} further noted that some bright AR loops are rooted in mixed polarity areas, while \cite{2017ApJ...843L..20T} suggested that interaction of opposite polarity fields may supply additional energy, in excess of that generated by loop braiding. Although the idea that the mixed polarity fields may play a role in coronal heating has been discussed for a while, the novelty of our study is that we were able to link a single flux emergence event to an isolated loop heating episode and to study the structure of the vector field associated with the emerging element. Recently \cite{2018ApJ...857...48G} used IRIS data to show that such magnetic cancellations may produce a clear signature of heating in the upper atmosphere, while \cite{2018ApJ...862L..24P} estimated parameters of three-dimensional reconnection driven by photospheric flux cancellation.

\cite{2003ApJ...593..549F} proposed that heating of quiet Sun coronal loops may be fueled by ``explosions'' of granule-scale sheared magnetic bipoles emerging at the edge of network flux concentrations. \cite{0004-637X-601-1-530} compared UV TRACE and SXT Yohkoh loops and concluded that cooler UV loops were mainly heated at their footpoints. \cite{2007ApJ...659.1673A} argued that heating occurs in the TR and the chromosphere and is due to photospheric rather than coronal magnetic complexity. \cite{2012ApJ...760...82S} studied magnetic field evolution at the footpoints of two loops using SDO/HMI data. High-speed outflows were observed in one loop where HMI data showed strong presence of mixed polarity fields at its footpoints. Another analyzed loop did not exhibit any detectable outflows but remained visible during nearly four hours, which significantly exceeds the estimated 65~min cooling time for a 200~Mm long loop \citep{0004-637X-720-2-1380}. What was the source of energy that ensured the loop visibility over a several hour period? Similarly to the case considered here, these authors did report a gradual increase of positive polarity flux at the loop footpoint, which may indicate varying magnetic fields. Reviewing the published data we also found that this loop was rooted at the edge of a network flux cluster, so that the scenario suggested by \cite{2003ApJ...593..549F} could be realized there. This is also similar to the two cases presented here. It is well known even since the pre-Hinode era that there are strong small-scale magnetic fields present in the granulation and associated with clusters of photospheric bright points \citep[e.g.,][and references therein]{Bart_2002,lites_2008}. These fields could carry enough energy for coronal heating as reported in \cite{2041-8205-810-2-L16}. \cite{2014ApJ...789..132R} arrived to the same conclusion using data from numerical simulations. Using line-of-sight data from IMeX instrument \citep{IMEX_2011} on the SUNRISE balloon \citep{Sunrise2010}, \cite{2017ApJS..229....4C} found small-scale mixed polarity magnetic fields at the footpoints of studied coronal loops and argued that flux cancellation and reconnection low in the solar atmosphere drive mass and energy flows along the loops. 

One of the loops studied here was rooted in a sunspot umbra away from a LB. It was gradually filled with plasma starting from the remote plage footpoint. \cite{2016A&A...587A..20C} discussed a case of a bright coronal loop with strong supersonic downflows rooted in a sunspot without LBs. We also note a difference between our and \cite{2016A&A...587A..20C} cases: while these authors were able to measure physical properties of the loop plasma using IRIS data, our loop was not visible in IRIS spectral lines, which indicates that its was generally hotter than the \cite{2016A&A...587A..20C} loop. Earlier, \cite{2015A&A...582A.116S} reported a case when a loop with supersonic downflows was not detectable in chromospheric lines, suggesting that sunspot rooted loops may exhibit various temperature and flow modes that sill need to be understood. \cite{2016A&A...587A..20C} speculated that a siphon flow generated by asymmetric heating at the other (unobserved in this case) footpoint may be the cause. We were able to trace the loop and locate its remote footpoint, which allowed us to identify small-scale magnetic activity in the photosphere. Based on HMI measurements and NIRIS data for Loop I we argue that small-scale ($\sim$1~Mm) flux emergence and cancellations have likely caused plasma injection into the loop via heating and evaporation, thus representing the asymmetric heating needed to drive siphon flows discussed in \cite{2016A&A...587A..20C}. 

Finally, the remote footpoints of both loops studies here were located at the edge of a plage region and, according to HMI data, they where rooted at or near small flux concentrations. It is known that small clusters of plage fields are associated with type II spicules \citep{2007PASJ...59S.655D}, and it was later argued \citep{bart_roots, 2011Sci...331...55D} that they may contribute to coronal heating as well. \cite{2013ApJ...767...17Y} analyzed NIRIS magnetic field data associated with a cluster of photospheric BPs and spicules of type II and reported that opposite polarity fields constantly appear in very close proximity to the cluster and the episodes of new flux emergence are connected to enhanced production of type II spicules.\cite{2017Sci...356.1269M} simulations of type II spicules further emphasized importance of emerging small-scale magnetic fields for their origin. Therefore, it is likely that magnetic reconnection and type II spicules are at the origin of warm coronal loops and further high resolution observations of loop footpoints in the photosphere and chromosphere are needed to shed more light on their structure.

The present study as well as earlier studies further show that on spatial scales below 2~Mm magnetic fields are still dynamic and complex, which may directly manifest itself in the solar corona through chromospheric and TR heating of coronal loops. X-ray flares in quiet Sun areas have recently been detected with Nuclear Spectroscopic Telescope ARray \citep[NuSTAR,][]{2041-8205-856-2-L32,2018ApJ...864....5M}, which once again emphasizes that the magnetic structures hidden below the resolution of modern instrumentation have to be taken into account when considering mechanisms of coronal heating.

\acknowledgments

SDO is a mission for NASA's Living With a Star (LWS) program. BBSO operation is supported by NJIT and US NSF AGS-1821294 grants. GST operation is partly supported by the Korea Astronomy and Space Science Institute (KASI), Seoul National University, and by strategic priority research program of CAS with Grant No. XDB09000000. This work is part of Mrs. S. {S}ahin's Master thesis and it was supported by Project 117F145 awarded by the Scientific and Technological Research Council of Turkey. VYu acknowledges support from AFOSR FA9550-15-1-0322 and NSF AST-1614457 grants. PK effort was supported by NASA Postdoctoral Program at the Goddard Space Flight Center, administered by the Universities Space Research Association through a contract with NASA. We thank Drs. Valentina Abramenko and Nai-Hwa Chen for valuable discussions and assistance. 


\bibliographystyle{aasjournal}
\bibliography{refs,my_refs,loops,sunspots}

\begin{thebibliography}{}
\expandafter\ifx\csname natexlab\endcsname\relax\def\natexlab#1{#1}\fi
\providecommand{\url}[1]{\href{#1}{#1}}
\providecommand{\dodoi}[1]{doi:~\href{http://doi.org/#1}{\nolinkurl{#1}}}
\providecommand{\doeprint}[1]{\href{http://ascl.net/#1}{\nolinkurl{http://ascl.net/#1}}}
\providecommand{\doarXiv}[1]{\href{https://arxiv.org/abs/#1}{\nolinkurl{https://arxiv.org/abs/#1}}}

\bibitem[{{Abramenko} {et~al.}(2009){Abramenko}, {Yurchyshyn}, \&
  {Watanabe}}]{ch_parameters}
{Abramenko}, V.~I., {Yurchyshyn}, V., \& {Watanabe}, H. 2009, \solphys, 260,
  43, \dodoi{10.1007/s11207-009-9433-7}

\bibitem[{{Abramenko} {et~al.}(2011){Abramenko}, {Yurchyshyn}, \&
  {Goode}}]{Abramenko_2011}
{Abramenko}, V.~I., {Yurchyshyn}, V.~B., \& {Goode}, P.~R. 2011, ArXiv
  e-prints.
\newblock \doarXiv{1112.2750}

\bibitem[{{Abramenko} {et~al.}(2013){Abramenko}, {Zank}, {Dosch}, {Yurchyshyn},
  {Goode}, {Ahn}, \& {Cao}}]{2013ApJ...773..167A}
{Abramenko}, V.~I., {Zank}, G.~P., {Dosch}, A., {et~al.} 2013, \apj, 773, 167,
  \dodoi{10.1088/0004-637X/773/2/167}

\bibitem[{{Aschwanden} {et~al.}(2013){Aschwanden}, {Boerner}, {Schrijver}, \&
  {Malanushenko}}]{2013SoPh..283....5A}
{Aschwanden}, M.~J., {Boerner}, P., {Schrijver}, C.~J., \& {Malanushenko}, A.
  2013, \solphys, 283, 5, \dodoi{10.1007/s11207-011-9876-5}

\bibitem[{Aschwanden \& Peter(2017)}]{0004-637X-840-1-4}
Aschwanden, M.~J., \& Peter, H. 2017, The Astrophysical Journal, 840, 4

\bibitem[{{Aschwanden} {et~al.}(2007){Aschwanden}, {Winebarger}, {Tsiklauri},
  \& {Peter}}]{2007ApJ...659.1673A}
{Aschwanden}, M.~J., {Winebarger}, A., {Tsiklauri}, D., \& {Peter}, H. 2007,
  \apj, 659, 1673, \dodoi{10.1086/513070}

\bibitem[{{Borrero} {et~al.}(2014){Borrero}, {Lites}, {Lagg}, {Rezaei}, \&
  {Rempel}}]{2014A&A...572A..54B}
{Borrero}, J.~M., {Lites}, B.~W., {Lagg}, A., {Rezaei}, R., \& {Rempel}, M.
  2014, \aap, 572, A54, \dodoi{10.1051/0004-6361/201424584}

\bibitem[{Brooks {et~al.}(2010)Brooks, Warren, \&
  Winebarger}]{0004-637X-720-2-1380}
Brooks, D.~H., Warren, H.~P., \& Winebarger, A.~R. 2010, The Astrophysical
  Journal, 720, 1380

\bibitem[{{Cao} {et~al.}(2012){Cao}, {Goode}, {Ahn}, {Gorceix}, {Schmidt}, \&
  {Lin}}]{niris}
{Cao}, W., {Goode}, P.~R., {Ahn}, K., {et~al.} 2012, in Astronomical Society of
  the Pacific Conference Series, Vol. 463, Second ATST-EAST Meeting: Magnetic
  Fields from the Photosphere to the Corona., ed. T.~R. {Rimmele},
  A.~{Tritschler}, F.~{W{\"o}ger}, M.~{Collados Vera}, H.~{Socas-Navarro},
  R.~{Schlichenmaier}, M.~{Carlsson}, T.~{Berger}, A.~{Cadavid}, P.~R.
  {Gilbert}, P.~R. {Goode}, \& M.~{Kn{\"o}lker}, 291

\bibitem[{{Cao} {et~al.}(2010){Cao}, {Gorceix}, {Coulter}, {Ahn}, {Rimmele}, \&
  {Goode}}]{Cao2010}
{Cao}, W., {Gorceix}, N., {Coulter}, R., {et~al.} 2010, Astronomische
  Nachrichten, 331, 636, \dodoi{10.1002/asna.201011390}

\bibitem[{{Chae} {et~al.}(2009){Chae}, {Park}, {Lites}, {Cheung}, {Magara},
  {Mariska}, \& {Reeves}}]{2009ASPC..415..101C}
{Chae}, J., {Park}, S., {Lites}, B., {et~al.} 2009, in The Second Hinode
  Science Meeting: Beyond Discovery-Toward Understanding, Vol. 415, 101

\bibitem[{{Chen} {et~al.}(2014){Chen}, {Peter}, {Bingert}, \&
  {Cheung}}]{2014A&A...564A..12C}
{Chen}, F., {Peter}, H., {Bingert}, S., \& {Cheung}, M.~C.~M. 2014, \aap, 564,
  A12, \dodoi{10.1051/0004-6361/201322859}

\bibitem[{{Chen} {et~al.}(2017){Chen}, {Rempel}, \&
  {Fan}}]{2017ApJ...846..149C}
{Chen}, F., {Rempel}, M., \& {Fan}, Y. 2017, \apj, 846, 149,
  \dodoi{10.3847/1538-4357/aa85a0}

\bibitem[{{Chitta} {et~al.}(2018){Chitta}, {Peter}, \&
  {Solanki}}]{2018A&A...615L...9C}
{Chitta}, L.~P., {Peter}, H., \& {Solanki}, S.~K. 2018, \aap, 615, L9,
  \dodoi{10.1051/0004-6361/201833404}

\bibitem[{{Chitta} {et~al.}(2016){Chitta}, {Peter}, \&
  {Young}}]{2016A&A...587A..20C}
{Chitta}, L.~P., {Peter}, H., \& {Young}, P.~R. 2016, \aap, 587, A20,
  \dodoi{10.1051/0004-6361/201527340}

\bibitem[{{Chitta} {et~al.}(2017){Chitta}, {Peter}, {Solanki}, {Barthol},
  {Gandorfer}, {Gizon}, {Hirzberger}, {Riethm{\"u}ller}, {van Noort}, {Blanco
  Rodr{\'{\i}}guez}, {Del Toro Iniesta}, {Orozco Su{\'a}rez}, {Schmidt},
  {Mart{\'{\i}}nez Pillet}, \& {Kn{\"o}lker}}]{2017ApJS..229....4C}
{Chitta}, L.~P., {Peter}, H., {Solanki}, S.~K., {et~al.} 2017, \apjs, 229, 4,
  \dodoi{10.3847/1538-4365/229/1/4}

\bibitem[{{Cirtain} {et~al.}(2013){Cirtain}, {Golub}, {Winebarger}, {de
  Pontieu}, {Kobayashi}, {Moore}, {Walsh}, {Korreck}, {Weber}, {McCauley},
  {Title}, {Kuzin}, \& {Deforest}}]{2013Natur.493..501C}
{Cirtain}, J.~W., {Golub}, L., {Winebarger}, A.~R., {et~al.} 2013, \nat, 493,
  501, \dodoi{10.1038/nature11772}

\bibitem[{{Cranmer} \& {van Ballegooijen}(2010)}]{cranmer_2010}
{Cranmer}, S.~R., \& {van Ballegooijen}, A.~A. 2010, \apj, 720, 824,
  \dodoi{10.1088/0004-637X/720/1/824}

\bibitem[{{Crooker} {et~al.}(2002){Crooker}, {Gosling}, \&
  {Kahler}}]{2002JGRA..107.1028C}
{Crooker}, N.~U., {Gosling}, J.~T., \& {Kahler}, S.~W. 2002, Journal of
  Geophysical Research (Space Physics), 107, 1028, \dodoi{10.1029/2001JA000236}

\bibitem[{{De Pontieu}(2002)}]{Bart_2002}
{De Pontieu}, B. 2002, \apj, 569, 474, \dodoi{10.1086/339231}

\bibitem[{{De Pontieu} {et~al.}(2009){De Pontieu}, {McIntosh}, {Hansteen}, \&
  {Schrijver}}]{bart_roots}
{De Pontieu}, B., {McIntosh}, S.~W., {Hansteen}, V.~H., \& {Schrijver}, C.~J.
  2009, \apjl, 701, L1, \dodoi{10.1088/0004-637X/701/1/L1}

\bibitem[{{de Pontieu} {et~al.}(2007){de Pontieu}, {McIntosh}, {Hansteen},
  {Carlsson}, {Schrijver}, {Tarbell}, {Title}, {Shine}, {Suematsu}, {Tsuneta},
  {Katsukawa}, {Ichimoto}, {Shimizu}, \& {Nagata}}]{2007PASJ...59S.655D}
{de Pontieu}, B., {McIntosh}, S., {Hansteen}, V.~H., {et~al.} 2007, \pasj, 59,
  655.
\newblock \doarXiv{0710.2934}

\bibitem[{{De Pontieu} {et~al.}(2011){De Pontieu}, {McIntosh}, {Carlsson},
  {Hansteen}, {Tarbell}, {Boerner}, {Martinez-Sykora}, {Schrijver}, \&
  {Title}}]{2011Sci...331...55D}
{De Pontieu}, B., {McIntosh}, S.~W., {Carlsson}, M., {et~al.} 2011, Science,
  331, 55, \dodoi{10.1126/science.1197738}

\bibitem[{{Del Zanna}(2003)}]{2003A&A...406L...5D}
{Del Zanna}, G. 2003, \aap, 406, L5, \dodoi{10.1051/0004-6361:20030818}

\bibitem[{{Del Zanna} \& {Mason}(2003)}]{2003A&A...406.1089D}
{Del Zanna}, G., \& {Mason}, H.~E. 2003, \aap, 406, 1089,
  \dodoi{10.1051/0004-6361:20030791}

\bibitem[{{Falconer} {et~al.}(2003){Falconer}, {Moore}, {Porter}, \&
  {Hathaway}}]{2003ApJ...593..549F}
{Falconer}, D.~A., {Moore}, R.~L., {Porter}, J.~G., \& {Hathaway}, D.~H. 2003,
  \apj, 593, 549, \dodoi{10.1086/376359}

\bibitem[{{Foukal}(1976)}]{1976ApJ...210..575F}
{Foukal}, P.~V. 1976, \apj, 210, 575, \dodoi{10.1086/154862}

\bibitem[{{Go{\v{s}}i{\'c}} {et~al.}(2018){Go{\v{s}}i{\'c}}, {de la Cruz
  Rodr{\'\i}guez}, {De Pontieu}, {Bellot Rubio}, {Carlsson}, {Esteban Pozuelo},
  {Ortiz}, \& {Polito}}]{2018ApJ...857...48G}
{Go{\v{s}}i{\'c}}, M., {de la Cruz Rodr{\'\i}guez}, J., {De Pontieu}, B.,
  {et~al.} 2018, \apj, 857, 48, \dodoi{10.3847/1538-4357/aab1f0}

\bibitem[{{Gupta} {et~al.}(2015){Gupta}, {Tripathi}, \&
  {Mason}}]{2015ApJ...800..140G}
{Gupta}, G.~R., {Tripathi}, D., \& {Mason}, H.~E. 2015, \apj, 800, 140,
  \dodoi{10.1088/0004-637X/800/2/140}

\bibitem[{{Hara} {et~al.}(2008){Hara}, {Watanabe}, {Harra}, {Culhane}, {Young},
  {Mariska}, \& {Doschek}}]{2008ApJ...678L..67H}
{Hara}, H., {Watanabe}, T., {Harra}, L.~K., {et~al.} 2008, \apjl, 678, L67,
  \dodoi{10.1086/588252}

\bibitem[{{Harra} \& {Abramenko}(2012)}]{2012ApJ...759..104H}
{Harra}, L.~K., \& {Abramenko}, V.~I. 2012, \apj, 759, 104,
  \dodoi{10.1088/0004-637X/759/2/104}

\bibitem[{{Hayashi} {et~al.}(2015){Hayashi}, {Hoeksema}, {Liu}, {Bobra}, {Sun},
  \& {Norton}}]{hmidata2}
{Hayashi}, K., {Hoeksema}, J.~T., {Liu}, Y., {et~al.} 2015, \solphys, 290,
  1507, \dodoi{10.1007/s11207-015-0686-z}

\bibitem[{{Hoeksema} {et~al.}(2014){Hoeksema}, {Liu}, {Hayashi}, {Sun},
  {Schou}, {Couvidat}, {Norton}, {Bobra}, {Centeno}, {Leka}, {Barnes}, \&
  {Turmon}}]{hmidata}
{Hoeksema}, J.~T., {Liu}, Y., {Hayashi}, K., {et~al.} 2014, \solphys, 289,
  3483, \dodoi{10.1007/s11207-014-0516-8}

\bibitem[{Hong {et~al.}(2017)Hong, Yang, Wang, Ji, Ji, \&
  Cao}]{1674-4527-17-3-25}
Hong, Z.-X., Yang, X., Wang, Y., {et~al.} 2017, Research in Astronomy and
  Astrophysics, 17, 25

\bibitem[{{Ji} {et~al.}(2012){Ji}, {Cao}, \& {Goode}}]{2012ApJ...750L..25J}
{Ji}, H., {Cao}, W., \& {Goode}, P.~R. 2012, \apjl, 750, L25,
  \dodoi{10.1088/2041-8205/750/1/L25}

\bibitem[{{Katsukawa}(2007)}]{2007ASPC..369..287K}
{Katsukawa}, Y. 2007, in Astronomical Society of the Pacific Conference Series,
  Vol. 369, New Solar Physics with Solar-B Mission, ed. K.~{Shibata},
  S.~{Nagata}, \& T.~{Sakurai}, 287

\bibitem[{{Katsukawa} \& {Tsuneta}(2005)}]{2005ApJ...621..498K}
{Katsukawa}, Y., \& {Tsuneta}, S. 2005, \apj, 621, 498, \dodoi{10.1086/427488}

\bibitem[{{Klimchuk}(2006)}]{2006SoPh..234...41K}
{Klimchuk}, J.~A. 2006, \solphys, 234, 41, \dodoi{10.1007/s11207-006-0055-z}

\bibitem[{Kong {et~al.}(2018)Kong, Pan, Yan, Wang, \& Li}]{2041-8205-863-2-L22}
Kong, D.~F., Pan, G.~M., Yan, X.~L., Wang, J.~C., \& Li, Q.~L. 2018, The
  Astrophysical Journal Letters, 863, L22

\bibitem[{Kuhar {et~al.}(2018)Kuhar, Krucker, Glesener, Hannah, Grefenstette,
  Smith, Hudson, \& White}]{2041-8205-856-2-L32}
Kuhar, M., Krucker, S., Glesener, L., {et~al.} 2018, The Astrophysical Journal
  Letters, 856, L32

\bibitem[{{Kumar} {et~al.}(2013){Kumar}, {Innes}, \&
  {Inhester}}]{2013ApJ...779L...7K}
{Kumar}, P., {Innes}, D.~E., \& {Inhester}, B. 2013, \apj, 779, L7,
  \dodoi{10.1088/2041-8205/779/1/L7}

\bibitem[{{Kumar} {et~al.}(2015){Kumar}, {Nakariakov}, \&
  {Cho}}]{2015ApJ...804....4K}
{Kumar}, P., {Nakariakov}, V.~M., \& {Cho}, K.-S. 2015, \apj, 804, 4,
  \dodoi{10.1088/0004-637X/804/1/4}

\bibitem[{{Landi Degl'Innocenti} {et~al.}(1992){Landi Degl'Innocenti},
  {Sanchez}, {Collados}, \& {Vazquez}}]{1992soti.book...71L}
{Landi Degl'Innocenti}, E., {Sanchez}, F., {Collados}, M., \& {Vazquez}, M.
  1992, Magnetic field measurements., 71

\bibitem[{{Lemen} {et~al.}(2012){Lemen}, {Title}, {Akin}, {Boerner}, {Chou},
  {Drake}, {Duncan}, {Edwards}, {Friedlaender}, {Heyman}, {Hurlburt}, {Katz},
  {Kushner}, {Levay}, {Lindgren}, {Mathur}, {McFeaters}, {Mitchell}, {Rehse},
  {Schrijver}, {Springer}, {Stern}, {Tarbell}, {Wuelser}, {Wolfson}, {Yanari},
  {Bookbinder}, {Cheimets}, {Caldwell}, {Deluca}, {Gates}, {Golub}, {Park},
  {Podgorski}, {Bush}, {Scherrer}, {Gummin}, {Smith}, {Auker}, {Jerram},
  {Pool}, {Soufli}, {Windt}, {Beardsley}, {Clapp}, {Lang}, \& {Waltham}}]{aia}
{Lemen}, J.~R., {Title}, A.~M., {Akin}, D.~J., {et~al.} 2012, \solphys, 275,
  17, \dodoi{10.1007/s11207-011-9776-8}

\bibitem[{{Lenz}(1999)}]{1999ApJ...517..497L}
{Lenz}, D.~D. 1999, \apj, 517, 497, \dodoi{10.1086/307177}

\bibitem[{{Lites} {et~al.}(1998){Lites}, {Skumanich}, \& {Martinez
  Pillet}}]{1998A&A...333.1053L}
{Lites}, B.~W., {Skumanich}, A., \& {Martinez Pillet}, V. 1998, \aap, 333, 1053

\bibitem[{{Lites} {et~al.}(2008){Lites}, {Kubo}, {Socas-Navarro}, {Berger},
  {Frank}, {Shine}, {Tarbell}, {Title}, {Ichimoto}, {Katsukawa}, {Tsuneta},
  {Suematsu}, {Shimizu}, \& {Nagata}}]{lites_2008}
{Lites}, B.~W., {Kubo}, M., {Socas-Navarro}, H., {et~al.} 2008, \apj, 672,
  1237, \dodoi{10.1086/522922}

\bibitem[{{Marsch} {et~al.}(2004){Marsch}, {Wiegelmann}, \&
  {Xia}}]{2004A&A...428..629M}
{Marsch}, E., {Wiegelmann}, T., \& {Xia}, L.~D. 2004, \aap, 428, 629,
  \dodoi{10.1051/0004-6361:20041060}

\bibitem[{{Marsh} {et~al.}(2018){Marsh}, {Smith}, {Glesener}, {Klimchuk},
  {Bradshaw}, {Vievering}, {Hannah}, {Christe}, {Ishikawa}, \&
  {Krucker}}]{2018ApJ...864....5M}
{Marsh}, A.~J., {Smith}, D.~M., {Glesener}, L., {et~al.} 2018, \apj, 864, 5,
  \dodoi{10.3847/1538-4357/aad380}

\bibitem[{Mart{\'i}nez-Pillet {et~al.}(2011)Mart{\'i}nez-Pillet, del
  Toro~Iniesta, {\'A}lvarez-Herrero, Domingo, Bonet,
  Gonz{\'a}lezÂ Fern{\'a}ndez, L{\'o}pezÂ Jim{\'e}nez, Pastor,
  GasentÂ Blesa, Mellado, Piqueras, Aparicio, Balaguer, Ballesteros,
  Belenguer, BellotÂ Rubio, Berkefeld, Collados, Deutsch, Feller, Girela,
  Grauf, Heredero, Herranz, Jer{\'o}nimo, Laguna, Meller, Men{\'e}ndez,
  Morales, OrozcoÂ Su{\'a}rez, Ramos, Reina, Ramos, Rodr{\'i}guez,
  S{\'a}nchez, Uribe-Patarroyo, Barthol, Gandorfer, Knoelker, Schmidt, Solanki,
  \& VargasÂ Dom{\'i}nguez}]{IMEX_2011}
Mart{\'i}nez-Pillet, V., del Toro~Iniesta, J.~C., {\'A}lvarez-Herrero, A.,
  {et~al.} 2011, Solar Physics, 268, 57, \dodoi{10.1007/s11207-010-9644-y}

\bibitem[{{Mart{\'\i}nez-Sykora} {et~al.}(2017){Mart{\'\i}nez-Sykora}, {De
  Pontieu}, {Hansteen}, {Rouppe van der Voort}, {Carlsson}, \&
  {Pereira}}]{2017Sci...356.1269M}
{Mart{\'\i}nez-Sykora}, J., {De Pontieu}, B., {Hansteen}, V.~H., {et~al.} 2017,
  Science, 356, 1269, \dodoi{10.1126/science.aah5412}

\bibitem[{{McIntosh} {et~al.}(1981){McIntosh}, {Cram}, \&
  {Thomas}}]{1981phss.conf....7M}
{McIntosh}, P.~S., {Cram}, L.~E., \& {Thomas}, J.~H. 1981, in The Physics of
  Sunspots, 7--54

\bibitem[{{Nagata} {et~al.}(2003){Nagata}, {Hara}, {Kano}, {Kobayashi},
  {Sakao}, {Shimizu}, {Tsuneta}, {Yoshida}, \& {Gurman}}]{2003ApJ...590.1095N}
{Nagata}, S., {Hara}, H., {Kano}, R., {et~al.} 2003, \apj, 590, 1095,
  \dodoi{10.1086/375127}

\bibitem[{{Ofman} \& {Wang}(2008)}]{2008A&A...482L...9O}
{Ofman}, L., \& {Wang}, T.~J. 2008, \aap, 482, L9,
  \dodoi{10.1051/0004-6361:20079340}

\bibitem[{{Parker}(1988)}]{1988ApJ...330..474P}
{Parker}, E.~N. 1988, \apj, 330, 474, \dodoi{10.1086/166485}

\bibitem[{{Pesnell} {et~al.}(2012){Pesnell}, {Thompson}, \& {Chamberlin}}]{sdo}
{Pesnell}, W.~D., {Thompson}, B.~J., \& {Chamberlin}, P.~C. 2012, \solphys,
  275, 3, \dodoi{10.1007/s11207-011-9841-3}

\bibitem[{{Peter}(2015)}]{2015RSPTA.37350055P}
{Peter}, H. 2015, Philosophical Transactions of the Royal Society of London
  Series A, 373, 20150055, \dodoi{10.1098/rsta.2015.0055}

\bibitem[{{Pontin} {et~al.}(2017){Pontin}, {Janvier}, {Tiwari}, {Galsgaard},
  {Winebarger}, \& {Cirtain}}]{2017ApJ...837..108P}
{Pontin}, D.~I., {Janvier}, M., {Tiwari}, S.~K., {et~al.} 2017, \apj, 837, 108,
  \dodoi{10.3847/1538-4357/aa5ff9}

\bibitem[{{Priest} {et~al.}(2018){Priest}, {Chitta}, \&
  {Syntelis}}]{2018ApJ...862L..24P}
{Priest}, E.~R., {Chitta}, L.~P., \& {Syntelis}, P. 2018, \apj, 862, L24,
  \dodoi{10.3847/2041-8213/aad4fc}

\bibitem[{{Reale}(2010)}]{2010LRSP....7....5R}
{Reale}, F. 2010, Living Reviews in Solar Physics, 7, 5,
  \dodoi{10.12942/lrsp-2010-5}

\bibitem[{Reale(2014)}]{Reale2014}
Reale, F. 2014, Living Reviews in Solar Physics, 11, 4,
  \dodoi{10.12942/lrsp-2014-4}

\bibitem[{{Rempel}(2014)}]{2014ApJ...789..132R}
{Rempel}, M. 2014, \apj, 789, 132, \dodoi{10.1088/0004-637X/789/2/132}

\bibitem[{{Sadykov} {et~al.}(2015){Sadykov}, {Vargas Dominguez}, {Kosovichev},
  {Sharykin}, {Struminsky}, \& {Zimovets}}]{2015ApJ...805..167S}
{Sadykov}, V.~M., {Vargas Dominguez}, S., {Kosovichev}, A.~G., {et~al.} 2015,
  \apj, 805, 167, \dodoi{10.1088/0004-637X/805/2/167}

\bibitem[{{Sams} {et~al.}(1992){Sams}, {Golub}, \&
  {Weiss}}]{1992ApJ...399..313S}
{Sams}, III, B.~J., {Golub}, L., \& {Weiss}, N.~O. 1992, \apj, 399, 313,
  \dodoi{10.1086/171926}

\bibitem[{{Scherrer} {et~al.}(2012){Scherrer}, {Schou}, {Bush}, {Kosovichev},
  {Bogart}, {Hoeksema}, {Liu}, {Duvall}, {Zhao}, {Title}, {Schrijver},
  {Tarbell}, \& {Tomczyk}}]{hmi}
{Scherrer}, P.~H., {Schou}, J., {Bush}, R.~I., {et~al.} 2012, \solphys, 275,
  207, \dodoi{10.1007/s11207-011-9834-2}

\bibitem[{Schmieder {et~al.}(2004)Schmieder, Rust, Georgoulis, Demoulin, \&
  Bernasconi}]{0004-637X-601-1-530}
Schmieder, B., Rust, D.~M., Georgoulis, M.~K., Demoulin, P., \& Bernasconi,
  P.~N. 2004, The Astrophysical Journal, 601, 530

\bibitem[{{Schou} {et~al.}(2012){Schou}, {Scherrer}, {Bush}, {Wachter},
  {Couvidat}, {Rabello-Soares}, {Bogart}, {Hoeksema}, {Liu}, {Duvall}, {Akin},
  {Allard}, {Miles}, {Rairden}, {Shine}, {Tarbell}, {Title}, {Wolfson},
  {Elmore}, {Norton}, \& {Tomczyk}}]{hmi2}
{Schou}, J., {Scherrer}, P.~H., {Bush}, R.~I., {et~al.} 2012, \solphys, 275,
  229, \dodoi{10.1007/s11207-011-9842-2}

\bibitem[{{Shimojo} {et~al.}(2017){Shimojo}, {Hudson}, {White}, {Bastian}, \&
  {Iwai}}]{2017ApJ...841L...5S}
{Shimojo}, M., {Hudson}, H.~S., {White}, S.~M., {Bastian}, T.~S., \& {Iwai}, K.
  2017, \apjl, 841, L5, \dodoi{10.3847/2041-8213/aa70e3}

\bibitem[{Solanki {et~al.}(2010)Solanki, Barthol, Danilovic, Feller, Gandorfer,
  Hirzberger, RiethmÃ¼ller, SchÃ¼ssler, Bonet, Pillet, del Toro~Iniesta,
  Domingo, Palacios, KnÃ¶lker, GonzÃ¡lez, Berkefeld, Franz, Schmidt, \&
  Title}]{Sunrise2010}
Solanki, S.~K., Barthol, P., Danilovic, S., {et~al.} 2010, The Astrophysical
  Journal Letters, 723, L127

\bibitem[{Song {et~al.}(2015)Song, Chae, Park, Cho, Lim, Ahn, \&
  Cao}]{2041-8205-810-2-L16}
Song, D., Chae, J., Park, S., {et~al.} 2015, The Astrophysical Journal Letters,
  810, L16

\bibitem[{{Straus} {et~al.}(2015){Straus}, {Fleck}, \&
  {Andretta}}]{2015A&A...582A.116S}
{Straus}, T., {Fleck}, B., \& {Andretta}, V. 2015, \aap, 582,
  \dodoi{10.1051/0004-6361/201525805}

\bibitem[{{Su} {et~al.}(2012){Su}, {Liu}, {Shen}, {Liu}, \&
  {Mao}}]{2012ApJ...760...82S}
{Su}, J.~T., {Liu}, Y., {Shen}, Y.~D., {Liu}, S., \& {Mao}, X.~J. 2012, \apj,
  760, 82, \dodoi{10.1088/0004-637X/760/1/82}

\bibitem[{Tian {et~al.}(2018)Tian, Yurchyshyn, Peter, Solanki, Young, Ni, Cao,
  Ji, Zhu, Zhang, Samanta, Song, He, Wang, \& Chen}]{0004-637X-854-2-92}
Tian, H., Yurchyshyn, V., Peter, H., {et~al.} 2018, The Astrophysical Journal,
  854, 92

\bibitem[{{Tiwari} {et~al.}(2017){Tiwari}, {Thalmann}, {Panesar}, {Moore}, \&
  {Winebarger}}]{2017ApJ...843L..20T}
{Tiwari}, S.~K., {Thalmann}, J.~K., {Panesar}, N.~K., {Moore}, R.~L., \&
  {Winebarger}, A.~R. 2017, \apj, 843, L20, \dodoi{10.3847/2041-8213/aa794c}

\bibitem[{{Tripathi} {et~al.}(2009){Tripathi}, {Mason}, {Dwivedi}, {del Zanna},
  \& {Young}}]{2009ApJ...694.1256T}
{Tripathi}, D., {Mason}, H.~E., {Dwivedi}, B.~N., {del Zanna}, G., \& {Young},
  P.~R. 2009, \apj, 694, 1256, \dodoi{10.1088/0004-637X/694/2/1256}

\bibitem[{{Ugarte-Urra} {et~al.}(2009){Ugarte-Urra}, {Warren}, \&
  {Brooks}}]{2009ASPC..415..241U}
{Ugarte-Urra}, I., {Warren}, H.~P., \& {Brooks}, D.~H. 2009, in Astronomical
  Society of the Pacific Conference Series, Vol. 415, The Second Hinode Science
  Meeting: Beyond Discovery-Toward Understanding, ed. B.~{Lites}, M.~{Cheung},
  T.~{Magara}, J.~{Mariska}, \& K.~{Reeves}
  (http://aspbooks.org/custom/publications/paper/415-0241.html), 241.
\newblock \url{http://aspbooks.org/custom/publications/paper/415-0241.html}

\bibitem[{van Ballegooijen {et~al.}(2017)van Ballegooijen, Asgari-Targhi, \&
  Voss}]{0004-637X-849-1-46}
van Ballegooijen, A.~A., Asgari-Targhi, M., \& Voss, A. 2017, The Astrophysical
  Journal, 849, 46

\bibitem[{{van Driel-Gesztelyi} \& {Green}(2015)}]{2015LRSP...12....1V}
{van Driel-Gesztelyi}, L., \& {Green}, L.~M. 2015, Living Reviews in Solar
  Physics, 12, 1, \dodoi{10.1007/lrsp-2015-1}

\bibitem[{{Viall} \& {Klimchuk}(2011)}]{2011ApJ...738...24V}
{Viall}, N.~M., \& {Klimchuk}, J.~A. 2011, \apj, 738, 24,
  \dodoi{10.1088/0004-637X/738/1/24}

\bibitem[{Wang(2016)}]{2041-8205-820-1-L13}
Wang, Y.-M. 2016, The Astrophysical Journal Letters, 820, L13

\bibitem[{Webb \& Zirin(1981)}]{Webb1981}
Webb, D., \& Zirin, H. 1981, Solar Physics, 69, 99, \dodoi{10.1007/BF00151258}

\bibitem[{Wiegelmann {et~al.}(2010)Wiegelmann, Solanki, Borrero, Pillet, del
  Toro~Iniesta, Domingo, Bonet, Barthol, Gandorfer, Kn??lker, Schmidt, \&
  Title}]{2041-8205-723-2-L185}
Wiegelmann, T., Solanki, S.~K., Borrero, J.~M., {et~al.} 2010, The
  Astrophysical Journal Letters, 723, L185

\bibitem[{{Winebarger} {et~al.}(2014){Winebarger}, {Cirtain}, {Golub},
  {DeLuca}, {Savage}, {Alexander}, \& {Schuler}}]{2014ApJ...787L..10W}
{Winebarger}, A.~R., {Cirtain}, J., {Golub}, L., {et~al.} 2014, \apjl, 787,
  L10, \dodoi{10.1088/2041-8205/787/1/L10}

\bibitem[{{Winebarger} {et~al.}(2002){Winebarger}, {Warren}, {van
  Ballegooijen}, {DeLuca}, \& {Golub}}]{2002ApJ...567L..89W}
{Winebarger}, A.~R., {Warren}, H., {van Ballegooijen}, A., {DeLuca}, E.~E., \&
  {Golub}, L. 2002, \apjl, 567, L89, \dodoi{10.1086/339796}

\bibitem[{{W{\"o}ger} \& {von der L{\"u}he}(2007)}]{kisip_code}
{W{\"o}ger}, F., \& {von der L{\"u}he}, O. 2007, Appl. Opt., 46, 8015

\bibitem[{{Yang} {et~al.}(2018){Yang}, {Longcope}, {Ding}, \&
  {Guo}}]{2018arXiv180206206Y}
{Yang}, K.~E., {Longcope}, D.~W., {Ding}, M.~D., \& {Guo}, Y. 2018, ArXiv
  e-prints.
\newblock \doarXiv{1802.06206}

\bibitem[{{Yurchyshyn} {et~al.}(2013){Yurchyshyn}, {Abramenko}, \&
  {Goode}}]{2013ApJ...767...17Y}
{Yurchyshyn}, V., {Abramenko}, V., \& {Goode}, P. 2013, \apj, 767, 17,
  \dodoi{10.1088/0004-637X/767/1/17}

\bibitem[{{Yurchyshyn} {et~al.}(2015){Yurchyshyn}, {Abramenko}, \&
  {Kilcik}}]{2015ApJ...798..136Y}
{Yurchyshyn}, V., {Abramenko}, V., \& {Kilcik}, A. 2015, \apj, 798, 136,
  \dodoi{10.1088/0004-637X/798/2/136}

\bibitem[{{Yurchyshyn} {et~al.}(2010){Yurchyshyn}, {Goode}, {Abramenko},
  {Chae}, {Cao}, {Andic}, \& {Ahn}}]{nst_jet_2010}
{Yurchyshyn}, V.~B., {Goode}, P.~R., {Abramenko}, V.~I., {et~al.} 2010, \apj,
  722, 1970, \dodoi{10.1088/0004-637X/722/2/1970}

\bibitem[{{Zeng} {et~al.}(2013){Zeng}, {Cao}, \& {Ji}}]{2013ApJ...769L..33Z}
{Zeng}, Z., {Cao}, W., \& {Ji}, H. 2013, \apjl, 769, L33,
  \dodoi{10.1088/2041-8205/769/2/L33}

\bibitem[{{Zhang} {et~al.}(2014){Zhang}, {Gorceix}, {Schmidt}, {Goode}, {Cao},
  {Rimmele}, \& {Coulter}}]{2014SPIE.9148E..50Z}
{Zhang}, X., {Gorceix}, N., {Schmidt}, D., {et~al.} 2014, in \procspie, Vol.
  9148, Adaptive Optics Systems IV, 914850

\end{thebibliography}



\end{document}